\documentclass[12pt]{article}

\input epsf
\usepackage{amssymb,wrapfig}
\usepackage[matrix,arrow,curve]{xy}
\usepackage{cite}

\makeatletter
\@addtoreset{equation}{section}
\makeatother

\def\nn {{\cal N}}
\def\rr {{\Bbb R}}
\def\cc {{\Bbb C}}

\def\zz {{\Bbb Z}}
\def\del {\partial}
\def\cy {Calabi--Yau}
\def\ka {K\"ahler}
\def\del {\partial}

\makeatletter
\@addtoreset{equation}{section}
\makeatother

\newcommand{\nc}{\newcommand}
\nc{\eps}{\epsilon}

\begin{document}

    \begin{titlepage}
    \begin{center}

                                    \hfill   EFI-07-22\\
                                    \hfill   SU-ITP-07/12
    \vskip .3in \noindent


    {\Large \bf{Generalized K\"ahler Potentials from Supergravity}}

    \bigskip

     Nick Halmagyi$^1$ and Alessandro Tomasiello$^2$\\

    \bigskip
    $^1$Enrico Fermi Institute, University of Chicago, Chicago, IL
    60637, USA \\
    $^2$ITP, Stanford University, Stanford, CA 94305-4060, USA


    \vskip .5in
    {\bf ABSTRACT }
    \vskip .1in
    \end{center}

    \noindent
We consider supersymmetric $\nn=2$ solutions with
non--vanishing NS three--form. Building on worldsheet results, we reduce the problem to a single generalized Monge--Amp\`ere equation
on the generalized \ka\ potential $K$ recently interpreted geometrically by Lindstr\"om,
Ro\v cek, Von Unge and Zabzine.
One input in the procedure is a holomorphic
function $w$ that can be thought of as the effective superpotential
for a D3 brane probe. The procedure is hence likely to be useful for finding gravity duals to
field theories with non--vanishing abelian superpotential,
such as Leigh--Strassler theories.
We indeed show
that a purely NS precursor of the Lunin--Maldacena dual to the $\beta$--deformed $\nn=4$
super--Yang--Mills falls in our class.

    \vfill
    \eject

    \tableofcontents

    \end{titlepage}

\section{Introduction} 
\label{sec:intro}

Supersymmetric supergravity solutions with flux have recently
started revealing their mathematical underpinnings. The concept
of generalized complex geometry \cite{hitchin-gcy,gualtieri} has clarified
for example the structure of $\nn=1$ solutions with RR and NS field--strengths \cite{gmpt2} and of $\nn=2$ solutions with NS flux \cite{gualtieri,jeschek-witt}\footnote{In both cases, the results
concern type II supergravity, which is the one of interest in this paper.}.

The reason for these mathematical structures is likely to have its origin in the worldsheet action of the string. This is actually
already manifest \cite{gualtieri} in the case with only NS flux, the only one for which we currently have worldsheet control. In that case, generalized complex geometry has recently helped show  \cite{lindstrom-rocek-vonunge-zabzine} that the most general model
with $(2,2)$ supersymmetry has an off--shell supersymmetric action.
A by--product of the proof is that there exists locally a ``generalized \ka\ potential'' $K$ for any such model.
This function $K$ generalizes the familiar \ka\ potential for the case without flux.

A $(2,2)$ model need not have an $\nn=2$ supergravity vacuum as a
target, for the same reason that a \ka\ manifold need not be \cy.
In the case without flux, $K$ has to satisfy $\det(\del_i\bar\del_{\bar j} K)={\rm const}$ for the target to be \cy. This is sometimes called
Monge--Amp\`ere equation.

We will see here something similar
for the most general $\nn=2$ background in which NS flux
is also present. Namely, the generalized \ka\ potential $K$ has to
satisfy a single differential equation, presented below in (\ref{eq:det}), in order for the manifold to be an $\nn=2$ supergravity vacuum with NS three--form.\footnote{That a single equation should
be sufficient was first suggested to us by M.~Zabzine.}

We obtain this from supergravity, using the methods of generalized
complex geometry, and this reproduces the one--loop computation in \cite{grisaru-massar-sevrin-troost}. It also gives some new geometrical
insight for the potential $K$; for example, we see that
$K$ still appears in an expression $\del \bar\del K $, see (\ref{eq:pureK}).

On the way to showing this, we obtain some results of more general use. Generalized complex geometry approaches the supersymmetry
problem via a {\it compatible pure pair} of two differential forms $\Phi_\pm$, satisfying certain
algebraic constraints reviewed in section \ref{sec:review}. The
$\Phi_\pm$ also determine a metric and a $B$--field, so that in
this paper we never write down the metric explicitly.
Supersymmetry is then equivalent to simple--looking differential
equations on the forms $\Phi_\pm$ (see (\ref{eq:N=2}) and (\ref{eq:t-reform}) below). We obtain in (\ref{eq:Bdielpuretilde}) a simple expression for the {\it generic} solution to the algebraic constraints for $\Phi_\pm$. This is a massage of an earlier computation \cite{minasian-petrini-zaffaroni}; we feel that the simplicity of 
(\ref{eq:Bdielpuretilde})
will allow to find RR solutions more easily, and to recognize more promptly their geometrical features. (For example, in the NS case (\ref{eq:Bdielpuretilde}) lets us recognize some similarities with
four--dimensional studies in \cite{hitchin-gcy,apostolov-gauduchon-grantcharov}.)

The specialization of the result (\ref{eq:Bdielpuretilde}) to NS $\nn=2$ vacua reads (\ref{eq:pureK}) and leads to the generalized Monge--Amp\`ere we mentioned earlier. We stress again that these $\Phi_\pm$, for a $K$ that satisfies (\ref{eq:det}), lead to a metric and $B$--field that satisfy the condition for an $\nn=2$ vacuum.

The other input in this construction is a holomorphic
function $w$. One can see using \cite{martucci} that $w$ is the effective four--dimensional superpotential for a D3 brane sitting
at the point in the internal manifold. This suggests an interpretation
for the $\nn=2$ vacua described in (\ref{eq:pureK}). Namely, by adding a large number of D3 branes to an NS background characterized by a function $w$, one would expect to obtain the gravity dual for a theory
with a nonabelian version of $w$.

One such solution already exists in the literature \cite{lunin-maldacena}. Although the main point of that paper
is the gravity dual to the so--called $\beta$--deformation of $\nn=4$
super--Yang--Mills, it also presents a purely NS analogue of that solution. We show in section \ref{sub:lm_pot} that this NS solution
is indeed in the form (\ref{eq:pureK}) for an appropriate choice
of coordinates and of $K$, (\ref{eq:Klm}).

For backgrounds with RR flux, in addition, there has also been
recent progress relating the spacetime structure of generalized
complex geometry and a worldsheet formalism \cite{Linch:2006ig}.
The authors study ${\rm SU}(3)$-structure deformations of
Calabi-Yau backgrounds to first order in RR and NS flux using
Berkovits' hybrid formalism \cite{Berkovits:1994wr}. They find
that the physical states must be modified and the constraints
imposed by worldsheet $(2,2)$ supersymmetry are equivalent to the
spacetime supersymmetry equations in the form derived in
\cite{gmpt2}. These calculations support the notion that
generalized complex geometry is the natural framework in which
to make contact with perturbative string theory.


\section{Pure spinor pairs and vacua} 
\label{sec:review}

In this section we will quickly review the definition of a compatible
pure spinor pair and their uses in reformulating geometrically
the supersymmetry conditions for vacua. For more thorough introductions
to generalized complex geometry and its applications see \cite{hitchin-gcy,gualtieri,gmpt3}.

First of all we will need an internal product in the space of forms
(due to Chevalley):
    \begin{equation}
     \label{eq:defMukai}
    ( A , B) \,\mathrm{vol} \equiv \Big(A\wedge
    \lambda(B)\Big)_{\rm top} \ , \qquad
    \lambda(C_k)=(-)^{[\frac k2]} C_k \ ,
    \end{equation}
where $k$ denotes the degree of the form. This pairing is antisymmetric
in six dimensions, and it is invariant under the action of
${\rm O}(6,6)$ on forms (for more details on this action see for
example \cite[Section 2.1]{t-reform}).

A {\it pure spinor} $\Phi$ is a differential form (of mixed degree;
thus a section of $\oplus_k \Lambda^k T^*$)
\begin{enumerate}
    \item whose annihilator in
    $(T\oplus T^*)\otimes \cc$ has complex dimension 6.
    \item such that $(\bar\Phi,\Phi)\neq 0$ everywhere on the manifold.
\end{enumerate}
 Its {\it type} is the smallest degree occurring in the form.
Thus $e^{iJ}$ (which is pure, if $J^3$ is a volume form, because annihilated by ${\iota_{\del_m} + i J_{mn} dx^n\wedge \ , \forall m}$)
has type 0. A decomposable three--form $\Omega=\xi_1\wedge\xi_2 \wedge \xi_3$ (which is pure, if $\Omega\wedge\bar \Omega$ is never zero, because it is annihilated by $\xi_i\wedge$ and their dual vector fields) has type 3. In fact, every pure spinor of type $k$ \cite[Prop.~2.24]{gualtieri} can be written locally as
\begin{equation}\label{eq:exp}
    \Phi= \xi_1\wedge \ldots \wedge \xi_k\wedge e^\alpha\
\end{equation}
for some one--forms $\xi_i$ and two--form $\alpha$.

A pair of pure spinors $\Phi_\pm$ is said to be 
 {\it compatible} if
\begin{enumerate}
    \item   the condition
    \begin{equation}\label{eq:comp}
        (\Phi_-, X \Phi_+) = 0 = (\bar\Phi_-, X \Phi_+) \ \ \forall X \in
        T\oplus T^*\
    \end{equation}
    is satisfied\footnote{For the equivalence of this condition
    with the one defined in \cite{gualtieri}, see \cite{t-reform}.}.
\item The two pure spinors $\Phi_\pm$ have equal norm,
    \begin{equation}
        \label{eq:eqnorm}
        (\bar\Phi_+,\Phi_+ )= (\bar \Phi_-,\Phi_-)\ .
    \end{equation}
\item
If the pair $\Phi_\pm$ satisfies the two previous conditions, it
defines a metric (see \cite{gualtieri,gmpt3} and below for details).
Then we also impose that the metric defined by $\Phi_\pm$  be positive definite.
\end{enumerate}

As an example of compatible pure pair, consider $\Phi_+=e^{-iJ}$
and $\Phi_-=\Omega$, the two pure spinors of types 0 and 3 we have
considered above. Condition 1.~above then reduces to $J\wedge \Omega=0$
(or, in other words, that $J$ is of type $(1,1)$ in the
almost complex structure $I$ defined by $\Omega$). Condition 2.~says
that $i\Omega \wedge\bar\Omega=\frac43 J^3$. Now, one can determine a metric
from the almost complex structure $I$ and from $J$ via $g_{mn}=J_{mp}I^p{}_n$. Condition 3.~then says that this metric should
be positive definite. These conditions, together, make $(J,\Omega)$
an ${\rm SU}(3)$ structure on the manifold.

More generally, the conditions for a compatible pure pair
determine an ${\rm SU}(3)\times {\rm SU}(3)$ structure on $T\oplus T^*$. By projecting the two ${\rm SU}(3)$ factors on $T$, one obtains
two ${\rm SU}(3)$ structures on $T$.
In the particular case we just saw, these two 
${\rm SU}(3)$ structures coincide.

Actually, two compatible pure spinors determine not just a metric $g$
but also a B--field, a normalization function (which is going to be a combination of the dilaton and the warping), and two spinors $\eta^{1,2}_+$ of positive chirality. (One can think of these two
spinors as the two ${\rm SU}(3)$ structures of the previous paragraph.)
Concretely, this means that $\Phi_\pm$ can be written
as
\begin{equation}\label{eq:otimes}
    \Phi_\pm = (e^B\wedge) \eta^1_+ \otimes \eta^{2\,\dagger}_\pm
\end{equation}
where the tensor product on the right hand side is to be understood
as a differential form using the Clifford map $\gamma^{m_1 \ldots m_k}\to dx^{m_1}\wedge\ldots\wedge dx^{m_k}$ (see \cite[Section 3.4]{gmpt3} for a proof of (\ref{eq:otimes}) and for more explanations on the Clifford map).
The spinors $\eta^{1,2}_+$ are in the spinor
bundle associated to the metric $g$. This metric also determines
a volume form ${\rm vol}$, which we use from now on in the definition
of the internal product (\ref{eq:defMukai}). Finally, (\ref{eq:otimes})
can also be read backwards: namely, it is also
true that given any $B$ and $\eta^{1,2}_+$, the right hand side
defines a compatible pure spinor pair. In particular, given such a
pair $\Phi_\pm$ and a two--form $B$, the new pair $e^B \wedge \Phi_\pm$ is also a compatible pure spinor pair; this is called $B$--transform.

The main reason to define a compatible pair is that the conditions
for supersymmetric backgrounds of the form
 ${\rm Minkowski}_4\times M_6$ can be reformulated completely in terms
of a compatible pair on $M_6$.

Namely, for an NS ${\cal N}=2$ solution \cite{gualtieri,jeschek-witt}, the metric has to be
a product $g_{10}=g_{{\rm Mink}_4}+ g_6 $, and there has to be a
compatible pair on $M_6$ with norm
\begin{equation}\label{eq:dilnorm}
    (\bar\Phi_\pm,\Phi_\pm )^{1/2}= e^{-\phi}
\end{equation}
such that
\begin{equation}\label{eq:N=2H0}
    (d+ H_0\wedge)\Phi_\pm=0 \ .
\end{equation}
Remembering that the pair $\Phi_\pm$ might already define
a non--zero $B$ (see (\ref{eq:otimes})), the total $H$ curvature
is $H=H_0+ dB$. One is free to change the pair to one that has
$B=0$, which would then be closed under $(d+H\wedge)$\footnote{$H$ is always assumed to be closed, so $(d+H\wedge)$ is a differential.};
one cannot
in general include all of $H_0 $ in the pair itself, unless one
promotes the global behavior of $\Phi_\pm$ from ordinary differential forms to something more `gerby'. This paper will
be mainly concerned with local properties (we will work on $\cc^3$);
it will be convenient, then, to have all of the $B$ field in the
pure spinor pair, and the relevant condition will simply be
\begin{equation}\label{eq:N=2}
    d \Phi_\pm =0\ .
\end{equation}

The condition for world--sheet $(2,2)$ supersymmetry is weaker  
than the condition (\ref{eq:N=2}) for target space $\nn=2$ supersymmetry. Namely, since we have
considered so far the case with no RR fluxes, we can consider the
sigma model with target space described by a compatible pure pair
$\Phi_\pm$ and ask under what conditions it has $(2,2)$
worldsheet supersymmetry. The answer is known as
{\it bihermitian} \cite{gates-hull-rocek} or {\it generalized \ka}
\cite{gualtieri} geometry, and it consists of the differential equations
\begin{equation}
    \label{eq:int}
    d \Phi_\pm = (\iota_{v_\pm} + \xi_\pm \wedge) \Phi_\pm
\end{equation}
for some vectors $v_\pm$ and one--forms $\xi_\pm$.\footnote{The generalized \ka\ condition is usually expressed in terms of generalized complex
structures ${\cal J}_\pm$, tensors that we review succinctly in  section \ref{sec:geom}. There is a slight loss in generality here, in that
for global reasons ${\cal J}_\pm$ might exist without $\Phi_\pm$; we are assuming that $c_1$ of two line bundles
are zero. This will not be important for our paper, that focuses on
local solutions anyway.} This condition is weaker than (\ref{eq:N=2}).
The reason is that the generalized \ka\ condition (\ref{eq:int}) guarantees  $(2,2)$ worldsheet supersymmetry, but not necessarily
conformal invariance, and hence (\ref{eq:int}) need not give rise to an $\nn=2$ vacuum. This is very familiar for models with ordinary \ka\ target spaces, that need not be \cy.

Finally, a similar result exists for backgrounds with non--vanishing RR
fields \cite{gmpt2,t-reform}.
For $\nn=1$ supersymmetry, the metric can now be relaxed to be a
warped product $g_{10}=e^{2A}g_{{\rm Mink}_4}+ g_6 $, for $A$ some function
on the internal $M_6$; the norm of the compatible pure spinor pair
now has to be
\begin{equation}\label{eq:Adilnorm}
    (\bar\Phi_\pm,\Phi_\pm )= e^{3A-\phi}
\end{equation}
and the differential equations are now (in IIB)
\begin{equation}\label{eq:t-reform}
    d \Phi_-=0 \ , \qquad  d (e^{-A}{\rm Re}\Phi_+)=0 \ ,   \qquad
    \delta= -8\,d d^{{\cal J }_-}(e^{-3A}{\rm Im}\Phi_+)
\end{equation}
where $\delta$ is the given magnetic source, and $d^{{\cal J}_-}$
is a differential defined from $\Phi_-$. (We have eliminated
the RR field $F$ from this equation, using the Bianchi identities.
For more details, see \cite{gmpt2,t-reform}.) Similarly to
(\ref{eq:N=2H0}), if one wants a non--trivial NS curvature $H_0$,
one can simply change $d\to (d+ H_0\wedge)$ in (\ref{eq:t-reform}).
We will see in the discussion after (\ref{eq:dPhi0}) that
the first equation in (\ref{eq:t-reform}) says that the manifold
should be {\it generalized complex.}

Both (\ref{eq:N=2}) and (\ref{eq:t-reform}) are reformulations
of the supersymmetry conditions. If one
also satisfies the Bianchi identities and equations of
motion for the fluxes, the remaining equations of motion follow \cite{lust-tsimpis,gauntlett-martelli-sparks-waldram-ads5-IIB}. For the NS flux $H$, we have assumed the Bianchi identity $dH=0$ throughout; the equation of motion $d *H=\ldots$ has
recently been shown in \cite{koerber-tsimpis} to follow from (\ref{eq:t-reform}). By taking a limit in which the RR fluxes go to
zero, (\ref{eq:t-reform}) reproduces (\ref{eq:N=2}) (with the amount of supersymmetry doubling in the process), so \cite{koerber-tsimpis}
also shows that the equation of motion for $H$ follows from (\ref{eq:N=2}). Turning to the RR fields, which are non--zero only
in (\ref{eq:t-reform}), their Bianchi identity can be easily shown
to follow from (\ref{eq:t-reform}) \cite{gmpt3}; as for their equations of motion, they have been used to eliminate the RR flux from (\ref{eq:t-reform}).


\section{Dielectric pure spinors} 
\label{sec:dielectric}
In this section we will show that a generic compatible pair of pure spinors can be
written, up to $B$--transform and common overall rescalings, as
\begin{equation}
    \label{eq:Bdielpuretilde}
    \fbox{$\begin{array}{c}\vspace{.2cm}
\Phi_+=
i\exp\left[ \frac12 z\wedge\bar z+i(\tilde\omega^1-\tilde\omega^2)
 \right]\\
\Phi_-= \tan(2\psi)z\wedge\exp\left[i(\tilde\omega^1+
\tilde\omega^2)
\right]
    \end{array}$}\ .
\end{equation}
Here, $z$ is a one--form,
 $\tilde\omega^{1,2}$ are two complex two--forms that satisfy
\begin{equation}\label{eq:omt}
    (\tilde\omega^1)^2=0=(\tilde \omega^2)^2 \ , \qquad {\rm Im} \tilde\omega^1=
    {\rm Im} \tilde \omega^2\equiv {\rm Im} \tilde \omega \ , \qquad ({\rm Im } \tilde\omega)^2 \neq 0\ \  {\rm everywhere}\  ,
\end{equation}
and
\begin{equation}
    \label{eq:psi}
    \tilde \omega^1 \wedge \tilde \omega^2+
    2 \sin^2(2 \psi)({\rm Im} \tilde \omega)^2=0\
\end{equation}
for $\psi$ a function (which is real, as follows from (\ref{eq:omt})).

With purely algebraic manipulations, (\ref{eq:Bdielpuretilde}) can be demonstrated up to a $B$-transform where $B$ is not necessarily closed. This will first be done in section \ref{sub:spinors} from ordinary Cliff(6) spinors and then in section \ref{sub:def} from the general definition of a compatible pair of pure spinors presented in section \ref{sec:review}.

The $B$--field associated to the pair (\ref{eq:Bdielpuretilde}) is
\begin{equation}
    \label{eq:B}
    B=2\sin^2(2\psi) {\rm Im} \tilde\omega\
\end{equation}
 and the norm is
\begin{equation}
    \label{eq:dil}
    (\bar\Phi_\pm,\Phi_\pm)^{1/2}= \frac 1{\cos(2 \psi)}\ .
\end{equation}
However, before we impose the differential constraints of 
supersymmetry (\ref{eq:int}), this is just a particularly nice choice. In section  \ref{sub:NS} we will impose the constraints (\ref{eq:int}) and derive (\ref{eq:Bdielpuretilde}) up to a $B$-transform with $dB=0$.

\subsection{From ordinary spinors} 
\label{sub:spinors}
We will now show how to obtain (\ref{eq:Bdielpuretilde}) if one
defines the pure spinors as bilinears of two internal spinors.

Let $\eta_+^{1,2}$ be two six--dimensional spinors of positive chirality
such that
\begin{equation}
    \label{eq:normeta}
    ||\eta_+^{1,2}||^2=1\ .
\end{equation}
By multiplying one of them by a phase if necessary, we can arrange for the scalar
\begin{equation}\label{eq:eta2real}
    \eta^{1\,\dagger}_+ \eta_+^2
\end{equation}
to be purely imaginary. Now define
\begin{equation}\label{eq:defetachi}
    \tilde\eta_+ =\frac12 (\eta_+^1 - i\eta^2_+)\ , \qquad
    \chi_+ =\frac12 (\eta_+^1 + i\eta^2_+)\ .
\end{equation}
In general, given two spinors of the same chirality, we can always
expand one in terms of the other; applying this to $\chi_+$
and $\tilde\eta_+$ we get\footnote{We define $\tilde\eta_-=(\tilde \eta_+)^*$; a similar convention will be used for all chiral spinors.}
\begin{equation}\label{eq:chiexp}
    \chi_+ = a \tilde\eta_+ + v\cdot\tilde\eta_-\
\end{equation}
for some complex function $a$ and vector $v$.
However, using (\ref{eq:normeta}) and (\ref{eq:eta2real}) we find
\begin{equation}
    \chi_+^\dagger \tilde\eta_+= -\frac i4(\eta_+^{1\,\dagger}\eta_+^2+
    \eta_+^{2\,\dagger}\eta_+^1)=-\frac i2 {\rm Re} (\eta_+^{1\,\dagger}\eta_+^2)=0\ .
\end{equation}
Comparing this with (\ref{eq:chiexp}) we see that $a=0$, or in other
words
\begin{equation}
    \chi_+ = v\cdot \tilde\eta_-\ .
\end{equation}
Going back to (\ref{eq:defetachi}), we obtain by sum and difference
\begin{equation}
    \eta^1_+= \tilde\eta_+ + v\cdot \tilde\eta_-\ , \qquad
    \eta^2_+= i(\tilde\eta_+ - v\cdot \tilde\eta_-)\ .
\end{equation}
We can now define $\eta_+=\tilde\eta_+/||\tilde\eta_+||$. Since
$\eta^{1\,\dagger}_- v \eta^2_+=0$, we have $||\eta^{1,2}_+||^2=||\tilde\eta_+||^2+||v\cdot\tilde\eta_-||^2$;
recalling (\ref{eq:normeta}), it follows that $||\tilde\eta_+||$ cannot be larger than 1. Hence we can define
\begin{equation}
    ||\tilde\eta_+||=\cos(\psi)\ ;
\end{equation}
from (\ref{eq:normeta}) then it also follows that $|v|=\tan(\psi)$.

Defining then $z=v/\sin(\psi)$, we have obtained
    \begin{equation}
        \label{eq:dielsp}
        \begin{array}{c}\vspace{.2cm}
    \eta_1^+= \cos(\psi) \eta_+ + \sin (\psi) z\cdot \eta_- \\
    \eta_2^+= i(\cos(\psi) \eta_+ - \sin(\psi) z\cdot \eta_-) \\
        \end{array}
    \end{equation}
where now $||\eta_+||=1=|z|$. We have shown that (\ref{eq:dielsp})
is the most general pair of spinors one can write, {\it up to} a phase
rotation for $\eta_+^2$ (that we fixed in (\ref{eq:eta2real})).

The spinors in (\ref{eq:dielsp}) are called dielectric spinors. In \cite{pilch-warner-gen-flow,pilch-warner-flow}
it was realized that certain holographic RG flows are in fact neat realizations of the dielectric or {\it Myers} effect \cite{myers}. In those solutions, the ten--dimensional spinors have the schematic form
\begin{equation}\label{eq:10dielsp}
\eps=\eps_1+i\eps_2\rightarrow \exp(i\psi\, \eps^{xy} \Gamma_{xy}*) \eps
\end{equation}
where $*\eps=\eps^*$.
These spinors  satisfy the projection conditions
\begin{equation}\label{eq:proj}
\eps= (\cos \psi\, + i \sin \psi\, \eps^{xy}\Gamma_{xy}*)\Gamma_{0123} \eps.
\end{equation}
Once we decompose the ten--dimensional $\eps^{1,2}$ in terms of Minkowski
and six--dimensional internal spinors as $\epsilon^{1,2}=\zeta_4\otimes \eta^{1,2}_+ + {\rm c. c.}$ as usual, (\ref{eq:10dielsp})
gives rise to (\ref{eq:dielsp}).

Physically, one interprets the projector (\ref{eq:proj}) to be the rotation of a D3 brane projector into a D5 brane projector; thus it is called a  {\it dielectric} projector.
We will now show that one can derive from the ``dielectric spinors''
in (\ref{eq:dielsp}) the expression for the ``dielectric pure spinors''
in (\ref{eq:Bdielpuretilde}).

From $ \eta^{1,2}$, one can define
a compatible pair $\Phi_\pm=\eta^1_+\otimes\eta^{2\,\dagger}_\pm$ just like in
(\ref{eq:otimes}). For (\ref{eq:dielsp}), $\Phi_\pm$ were
computed in \cite{minasian-petrini-zaffaroni}. We can repackage them
as follows:
    \begin{equation}
        \label{eq:dielpure}
        \begin{array}{c}\vspace{.2cm}
    \Phi_+= i\cos(2\psi)\exp\left[-\frac i{\cos(2\psi)}j+\frac12 z\wedge\bar z
    +\tan(2\psi){\rm Im} \omega\right]\\
    \Phi_-= \sin(2\psi)z \wedge
    \exp\left[\frac i{\sin(2\psi)}{\rm Re} \omega-
    \frac{\cos(2\psi)}{\sin(2\psi)}{\rm Im} \omega\right]
        \end{array}\ .
    \end{equation}
Here, $\omega$ and $j$ describe, together with $z$,
an SU(2) structure on $M_6$. It is inside the SU(3) structure
$(J=j+\frac i2 z\wedge\bar z$, $\Omega=\omega\wedge z)$ defined by $\eta_+$ through $8\eta_+\otimes \eta_+^\dagger\equiv e^{-iJ}$ and
$8i \eta_+\otimes \eta_-^\dagger \equiv  \Omega$.

Each of the $\eta^i$ alone also defines an SU(3) structure via
$8\eta^i_+ \otimes \eta_+^{i\,\dagger}\equiv e^{-i J^i}$ and $8i \eta^i_+ \otimes \eta_-^{i\,\dagger} \equiv \Omega^i$;
and again each of the two SU(3) structures defines, together with $z$, an SU(2) structure:
\begin{equation}\label{eq:Jj}
    \Omega^i=\omega^i \wedge z\ , \qquad
     J^i=j^i +\frac i2 z\wedge\bar z \ .
\end{equation}
     One can compute
\begin{equation}
        \label{eq:su3i}
        \begin{array}{c}\vspace{.2cm}
        \omega^{1,2}= \cos(2\psi){\rm Re} \omega\mp \sin(2\psi) j+i {\rm Im} \omega =
    {\rm Re}\Big(e^{\pm 2i\psi}({\rm Re} \omega+ i j)\Big)+ i {\rm Im} \omega\ ,\\
       j^{1,2}=\cos(2\psi) j\pm \sin(2\psi) {\rm Re} \omega=
    {\rm Im}\Big(e^{\pm 2i\psi}({\rm Re} \omega+ i j)\Big)\ .
        \end{array}
\end{equation}

Now we want to try and reexpress the pure spinors (\ref{eq:dielpure}) in terms of
$\omega^{1,2}$ in (\ref{eq:su3i}). We also have the freedom of
taking a $B$--transform
\begin{equation}
\Phi_\pm \to e^{-B}\wedge \Phi_\pm \ ;
\end{equation}
notice that so far (\ref{eq:dielpure})
have been defined by $\Phi_\pm=\eta^1_+\otimes\eta^{2\, \dagger}_\pm$, and
hence, comparing with (\ref{eq:otimes}), they define a zero $B$.

It so happens that the best choice for $B$ is such that
the exponent of $\Phi_+$ is purely imaginary:
\begin{equation}\label{eq:Bnont}
    B=\tan(2 \psi){\rm Im} \omega\ .
\end{equation}
In section \ref{sub:NS} we will see that the differential constraints of NS ${\cal N}=2$ backgrounds impose
this choice  (up to a {\it closed} $B$); it actually also
happens to make the expression for $\Phi_\pm$ more pleasant--looking:
\begin{equation}
        \label{eq:Bdielpure}
        \begin{array}{c}\vspace{.2cm}
    \Phi_+= i\cos(2\psi)\exp\left[\frac12 z\wedge\bar z+\frac i{\sin(4\psi)}(\omega^1-\omega^2)
     \right],\\
    \Phi_-= \sin(2\psi)z\wedge\exp\left[\frac i{\sin(4\psi)}(\omega^1+
    \omega^2)
    \right].
        \end{array}\
\end{equation}

So far (\ref{eq:Bdielpure}) have norm $(\bar\Phi_\pm,\Phi_\pm )= 1$,
since we have taken $||\eta||=1$ in (\ref{eq:dielsp}).
We can also rescale (\ref{eq:Bdielpure}),  to obtain, after
defining
\begin{equation}\label{eq:tom-sp}
\tilde\omega^a = \frac1{\sin(4 \psi)}   \omega^a\ ,
\end{equation}
the compatible pure spinor pair in (\ref{eq:Bdielpuretilde}),
with norm  given by (\ref{eq:dil}); the $B$ field in (\ref{eq:Bnont})
turns into (\ref{eq:B}).
In section \ref{sub:NS} we will see that this choice 
of normalization is forced on us in the case of NS $\nn=2$ backgrounds,
just as it was the case for the choice of B--field (as remarked
after (\ref{eq:Bnont})). One can check using (\ref{eq:su3i})
that the $\tilde \omega^a $ defined in (\ref{eq:tom-sp}) satisfy
(\ref{eq:omt}) and (\ref{eq:psi}).


\subsection{From the definition} 
\label{sub:def}

We will now also show how to obtain (\ref{eq:Bdielpuretilde}) from
the definition of compatible pure spinor pair.

First of all, a generic pure spinor pair in six dimensions
has types 0 and 1; namely, the form of lowest degree in $\Phi_+$ is
a zero--form, and in $\Phi_-$, a one--form.
We also know from (\ref{eq:exp}) that a pure spinor of
type 1 can always be written as
\begin{equation} \label{eq:Pm}
\Phi_-=\phi_1\wedge e^{\alpha}
\end{equation}
for some one--form $\phi_1$ and two--form $\alpha$. Similarly,
a pure spinor $\Phi_+$ of type 0 can be written as the 
exponential of a two--form. Without loss of generality we can take
\begin{equation} \label{eq:Pp}
\Phi_+=e^{\beta+ f \phi_1 \wedge \bar \phi_1},
\end{equation}
for some function $f$. One could also allow
for another function in front of the exponential; however, given
a compatible pair, even after taking into account (\ref{eq:eqnorm}),
one has the freedom to rescale both $\Phi_\pm$ by a function, and
we will fix this ambiguity by taking the zero--form in $\Phi_+$
to be just 1.

For $v=v_m dx^m$ a one-form, we denote the contraction $v_m(E^{-1})^{mn}\iota_{\del_n}$ by $v\llcorner$. Here as usual $E=g+B$. Now we use this contraction operation to decompose an arbitrary two-form $\omega$ as
\begin{equation} \label{eq:orthdec}
\omega=\omega'+\frac{\phi_1\wedge (\bar{\phi}_1\llcorner \omega)}{\bar{\phi}_1\llcorner \phi_1}+\frac{\bar{\phi}_1\wedge (\phi_1\llcorner \omega)}{\phi_1\llcorner \bar{\phi}_1} -\phi_1\wedge \bar{\phi}_1 \frac{\phi_1\llcorner \bar{\phi}_1\llcorner \omega}{(\bar{\phi}_1\llcorner \phi_1)(\phi_1\llcorner \bar{\phi}_1)}
\end{equation}
so that
\begin{equation}
\phi_1\llcorner \omega' =0,\ \ \ \bar{\phi}_1 \llcorner \omega'=0.
\end{equation}

To find out about the properties of $\phi_1 \llcorner$ and its
conjugate, one has to compute $g$ and $B$ from (\ref{eq:Pp}) and
(\ref{eq:Pm}). Actually, one can take a shortcut by using (\ref{eq:otimes}) to translate
the annihilators of $\Phi_\pm$ in terms of gamma matrices; for
more details see for example \cite[Sec.~3.4]{gmpt3}. This gives
the conditions
\begin{equation}
(\phi_1 \wedge+ \phi_1 \llcorner )\Phi_\pm=0\ , \qquad
    \begin{array}{c}
    (\phi_1 \wedge- \phi_1 \llcorner )\bar\Phi_+=0\ ,\\
    (\phi_1 \wedge- \phi_1 \llcorner )\Phi_-=0\ ,
    \end{array}
\end{equation}
and their conjugates. Imposing
this and applying the decomposition (\ref{eq:orthdec}) to the two--forms $\alpha,\beta$,
after some algebra we find that we can rewrite the pure spinors
(\ref{eq:Pp}) and (\ref{eq:Pm}) again as
\begin{equation}
    \begin{array}{l}\vspace{.2cm}
        \Phi_-=\phi_1 \wedge e^{\alpha'}\ , \\
\Phi_+=\exp\left(\beta' + f' \phi_1\wedge \bar{\phi}_1\right)\ ,
    \end{array}
\end{equation}
but this time in terms of new $\alpha'$, $\beta'$, $f'$ such that
 $\phi_1\llcorner$ and $\bar{\phi}_1\llcorner$ annihilate $\alpha'$ and $\beta'$, and the function $f'$ is real. From now on we will drop the
primes.

We now apply (\ref{eq:comp}) to $\Phi_\pm$. If $X$ is a one--form $\zeta$, we get
\begin{equation}
    \phi_1 \wedge \zeta \wedge (\beta-\alpha)^2=0=
    \bar\phi_1 \wedge \zeta \wedge (\beta-\bar\alpha)^2\
\end{equation}
 which implies
\begin{equation}\label{eq:comp2}
    (\beta-\alpha)^2=0= (\beta-\bar\alpha)^2\ .
\end{equation}
The case in which $X$ in (\ref{eq:comp}) is a vector does not give
any extra condition.

We can already see from (\ref{eq:comp2}) that $\beta-\alpha$ and
$\beta-\bar \alpha$ have the properties required by (\ref{eq:omt}).
However, the forms that appear in $\Phi_\pm$ are $\alpha$ and $\beta$.
To make contact between the two, recall once again that, given a pure spinor compatible pair, one can always produce another by
$\Phi_\pm\to e^B\wedge \Phi_\pm$ for $B$ any real two--form. Using this, we can choose to make $\beta$ purely imaginary. As in the previous subsection, this is just a choice at this point, but it
will be pointed out in the next subsection that it is actually
necessary for NS $\nn=2$ backgrounds.

Having made $\beta$ purely imaginary, we can define
\begin{equation}
    \alpha-\beta= 2 i \tilde \omega_2 \ , \qquad \alpha+\beta=2 i \tilde \omega_1 \ ,
\end{equation}
and by (\ref{eq:comp2}) we conclude (\ref{eq:omt}).
To summarize, so far we have obtained that the pure spinors
can be written, up to $B$--transform, as
\begin{equation}\label{eq:dieltemp}
    \Phi_+ = \exp[i(\tilde \omega^1 - \tilde \omega^2)+ f \phi_1 \wedge \bar \phi_1]\ , \qquad \Phi_- = \phi_1 \wedge \exp[i(\tilde \omega^1 + \tilde \omega^2)]\ .
\end{equation}

This takes care of the first condition for compatibility that we saw in section \ref{sec:review}, namely (\ref{eq:comp}). We now turn to (\ref{eq:eqnorm}). For that, notice first that (\ref{eq:omt}) (which we just derived from (\ref{eq:comp2})) implies $({\rm Re} \tilde \omega^1)^2=({\rm Re} \tilde \omega^2)^2=({\rm Im} \tilde \omega)^2$ and ${\rm Re} \tilde \omega^a \wedge {\rm Im} \tilde \omega=0 $.
The wedge product ${\rm Re} \tilde \omega^1 \wedge {\rm Re} \tilde \omega^2$, however, is not determined by this. Hence we can define the function $\psi$ by
\begin{equation}
    \label{eq:psi2}
     {\rm Re} \tilde \omega^1 \wedge {\rm Re} \tilde \omega^2+(1-2 \cos^2(2 \psi))({\rm Im} \tilde \omega)^2 =0\
\end{equation}
and then apply (\ref{eq:eqnorm}) to
(\ref{eq:dieltemp}). We get that
\begin{equation}
    f=\frac1{2 \tan^2(2 \psi)}\ .
\end{equation}
By taking now $z=\frac{\phi_1}{\tan(2 \psi)}$ one finally finds
(\ref{eq:Bdielpuretilde}). ((\ref{eq:psi2}) reduces then to (\ref{eq:psi}).)


\subsection{NS backgrounds} 
\label{sub:NS}

In deriving (\ref{eq:Bdielpuretilde}), we had to fix two ambiguities:
under $B$--transform ($\Phi_\pm\to e^B\wedge \Phi_\pm$) and rescaling
($\Phi_\pm \to f \Phi_\pm$). We will now show that the choices
we made are actually necessary in the case of NS backgrounds with
$\nn=2$ supersymmetry.

The differential equations are quite simple: they say that $\Phi_\pm$
are closed, (\ref{eq:N=2}). Let us focus on $\Phi_+$, and let us
go back to the expression for it given in (\ref{eq:dielpure})
\begin{equation}
\Phi_+= \cos(2\psi)\exp\left[-\frac i{\cos(2\psi)}j+\frac12 z\bar z
+\tan(2\psi){\rm Im} \omega\right]
\end{equation}
which has norm 1 and $B=0$.
(Both the norm and $B$ field do depend on what $\Phi_-$ is; they would
be different if we changed $\Phi_-$ in (\ref{eq:dielpure}).)

Now let us rescale and $B$--transform this $\Phi_+$ (supposing we also
do the same to $\Phi_-$), and impose $d(f e^{-B}\wedge\Phi_+)=0$. First
of all we see that $f=1/\cos(2 \psi)$. This explains
the rescaling made at the end of section \ref{sub:spinors},
to go from (\ref{eq:Bdielpure}) to (\ref{eq:Bdielpuretilde}), which
gives (\ref{eq:dil}). In particular, remembering (\ref{eq:dilnorm}),
we have
\begin{equation}\label{eq:dil2}
    e^\phi=\cos(2 \psi)\ .
\end{equation}

Then we also see that the exponent of $e^B\wedge \Phi_+$ should be closed, which means
\begin{equation}
    d\Big(B  +\frac i{\cos(2\psi)}j-\frac12 z\bar z
    -\tan(2\psi){\rm Im} \omega\Big)=0 \ .
\end{equation}
 The real part of this equation implies that $B=\tan(2 \psi){\rm Im}  \omega+ B_0$, where $B_0$ is closed.
In  (\ref{eq:B}) we took $B_0=0$, since the focus of this paper is
on local properties.

Now that we have justified the choices made to fix the ambiguities
in (\ref{eq:Bdielpuretilde}), we can also impose that they be closed.
For completeness, we write them here:
\begin{equation}\label{eq:dom}
    d(\tan(2\psi)z)=0 \ , \qquad z\wedge d(\tilde\omega^1+
    \tilde\omega^2)=0 \ , \qquad d\Big(\tilde\omega^1-\tilde\omega^2-
    \frac i2 z\wedge\bar z\Big)=0 \ .
\end{equation}


\subsection{Branes} 
\label{sub:branes}

The only reason we gave so far for being interested in the compatible
pair (\ref{eq:Bdielpuretilde}) is that it is the most general pair
of types 0 and 1, and hence the generic pure spinor pair.
A more compelling and physical reason to be interested 
in backgrounds of this type is that the
moduli space of D3 brane probes is partially lifted. This point was
originally made in \cite{martucci} but we repeat it here for the
reader's convenience.

The fact that the moduli space of D3 branes is lifted can be seen in various ways. The most straightforward is
to use the conditions in \cite{koerber,martucci-smyth,martucci}:
\begin{equation}\label{eq:branes}
    \iota^*[{\rm Re} \Phi_+]|_{\rm top}=0\ , \qquad
    \iota^*[(\iota_{\del_m }+g_{mn}dx^n\wedge)\Phi_-]|_{\rm top}=0 \ ,
\end{equation}
where $\iota: \,B\hookrightarrow M_6$ is the inclusion, and as usual
$\iota^*$ is the pull--back.
The symbol $|_{\rm top}$ means that one should keep the form
of highest degree on $B$.
These conditions
generalize, and are derived in the same way as, the ones for backgrounds without fluxes: see for example \cite{becker-becker-strominger,marino-minasian-moore-strominger}. Also, they reproduce physically the mathematical definition of generalized
complex submanifolds given in \cite{gualtieri}.

So, consider D3 brane--probes extended along Minkowski$_4$ and located at a point in $M_6$, in a background described by (\ref{eq:Bdielpuretilde}).
The first condition in (\ref{eq:branes})
is automatically satisfied, thanks to the $i$ in front of $\Phi_+$ in
(\ref{eq:Bdielpuretilde}). The second condition is satisfied only
at points where
\begin{equation}\label{eq:d3}
    \tan(2 \psi) z=0.
\end{equation}

Also, recall that, if (\ref{eq:Bdielpuretilde}) describe an $\nn=2$
solution, $\Phi_\pm$ must be closed. This implies (as we have seen already in (\ref{eq:dom})) that $d(\tan(2 \psi)z)=0$. Locally, this means that
\begin{equation}\label{eq:zw}
    \tan(2 \psi)z =dw
\end{equation}
 for some function $w$. In fact, one can go further
and argue \cite{martucci} that $w$ is nothing else than the four--dimensional superpotential (for a {\it single} brane probe). As a check, supersymmetric vacua for the effective four--dimensional theory
are critical points for $w$. At these points $dw=0$, which, remembering
(\ref{eq:zw}) and (\ref{eq:d3}), is precisely the condition for
the D3 brane to be supersymmetric.

This result is important for us: it tells us that the class of metrics
we are considering can be trivially adjusted so as to produce an
arbitrary assigned superpotential $w$ on the four--dimensional effective theory on D3 branes.

Another interesting case to consider is the case of a D7 wrapping
the submanifold $\{ w=w_0\}$. (We will see why this is interesting geometrically in section \ref{sec:geom}.) In \cite{mariotti} it has been shown for a few examples
of solutions (albeit with nonvanishing RR fields)
that this is a supersymmetric cycle.

In general, wrapping the submanifold $\{ w=w_0\}$ with the $B$--field
we gave in (\ref{eq:B}) (and hence with the pure spinors in
(\ref{eq:Bdielpuretilde})) does not satisfy the conditions (\ref{eq:branes}) for a supersymmetric brane. However, in section \ref{sec:lm} we will consider a solution for which a different
choice $B'=B+ d \lambda$ exists\footnote{This different choice is not gauge equivalent to $B$, because we are not transforming $A'= A -\lambda$ at the same time.}
 (originally considered in \cite{lunin-maldacena})
so that the supersymmetry conditions with that $B'$ are satisfied.

We will have more to say about this case in section \ref{sub:nc}.


\subsection{A four--dimensional analogue} 
\label{sub:4d}
It might be useful to notice how the results
obtained so far extend those obtained in \cite{apostolov-gauduchon-grantcharov,hitchin-delpezzo}.
We will also see in section \ref{sub:leaf}
that this situation is relevant for us because
of a certain foliation we will define in section \ref{sec:geom}.

The papers \cite{apostolov-gauduchon-grantcharov,hitchin-delpezzo} consider the generalized \ka\ condition (a weaker version of the $\nn=2$ condition, which we reviewed around equation (\ref{eq:int})). They point out that a generalized \ka\
manifold in four dimensions can be obtained by two two--forms $\tilde\omega^{1,2}_{(4)}$
\begin{equation}
\begin{array}{cc}\vspace{.2cm}
    (\tilde\omega^1_{(4)})^2=0=(\tilde\omega^2_{(4)})^2\ , \qquad {\rm Im } \tilde\omega^1_{(4)}=
    {\rm Im }\tilde\omega^2_{(4)}\equiv {\rm Im }\tilde\omega_{(4)}\ , \\
    ({\rm Im } \tilde\omega_{(4)})^2 \neq 0\ \  {\rm everywhere}\  , \qquad
    d \tilde\omega^{1,2}_{(4)}=0 \ .
\end{array}
    \label{eq:om4}
\end{equation}
The algebraic equations in (\ref{eq:om4}) are exactly like those
for $\tilde\omega^{1,2}$ in (\ref{eq:omt}). The requirement that $\tilde\omega^{1,2}_{(4)}$ should be closed is less exactly paralleled by (\ref{eq:dom}).

In fact, we can explain (\ref{eq:om4}) from pure
spinors, and explain the difference with (\ref{eq:dom}) (relevant
to the six--dimensional case) in the
process. One can consider the pair
\begin{equation}
    \label{eq:4dpair}
    \Phi_1^{(4)} = \cos(2 \psi) e^{i (\tilde\omega^1_{(4)}- \tilde\omega^2_{(4)}) }\ , \qquad
    \Phi_2^{(4)} = \sin(2 \psi) e^{i (\tilde\omega^1_{(4)}+ \tilde\omega^2_{(4)}) }\ ,
\end{equation}
which can be derived, as in section \ref{sub:spinors}, from four--dimensional spinors
\begin{equation}
        \label{eq:4ddielsp}
        \begin{array}{c}\vspace{.2cm}
    \eta^1_{(4)}= \cos(\phi) \eta_{(4)} + \sin(\psi)  \eta^*_{(4)} \ ,\\
    \eta^2_{(4)}= i(\cos(\phi) \eta_{(4)} - \sin(\psi) \eta^*_{(4)})\ . \\
        \end{array}
\end{equation}

The condition for a pair $\Phi_\pm$ to define a
generalized \ka\ structure is
(\ref{eq:int}); we will show in section \ref{sub:checking}
that applying (\ref{eq:int}) to a pure
spinor of type 0, $\Phi= f e^\alpha$, implies that the exponent
is closed, $d \alpha=0 $. Using this in (\ref{eq:4dpair}),
we obtain that $d (\omega^{1,2}_{(4)})=0$, which is the differential
equation in (\ref{eq:om4}).



\section{Geometrical interpretation} 
\label{sec:geom}

In this section, we will make some mathematical comments about the compatible pure spinor pair
(\ref{eq:Bdielpuretilde}), many of which will be needed in section \ref{sec:gk}. In section \ref{sub:purepoisson} we will 
spell out some relations between pure spinors and Poisson geometry,
and review
a splitting of the tangent bundle $T$ introduced in \cite{lindstrom-rocek-vonunge-zabzine}. This splitting will then 
be given a sigma--model interpretation in section \ref{sub:sigma},
again following \cite{lindstrom-rocek-vonunge-zabzine}. In section
\ref{sub:leaf} we will focus on one particular
summand in this decomposition of $T$. 
Section \ref{sub:nc} explains a possible spinoff in noncommutative
geometry.

\subsection{Pure spinors and Poisson geometry}
\label{sub:purepoisson}

The aim of this subsection is to show and explain the decomposition
(\ref{eq:decomp}) of the tangent bundle $T$.

We will start by recalling some mathematical definitions. A {\it Poisson} tensor $P^{mn}$ is a bivector (a section of
$\Lambda^2 T$)
such that
\begin{equation}\label{eq:Poisson}
    P^{[m|q}\del_q P^{|np]}=0 \ .
\end{equation}
In index--free notation, (\ref{eq:Poisson}) can be written as 
$[P,P]_{\rm NS}=0$, where $[\,,]_{\rm NS}$ is the  
Nijenhuis--Schoutens bracket on $\Lambda^k T$.
A {\it foliation} is a partition of a manifold $M_6$ in disjoint
connected sets (called {\it leaves}), such
that in every chart of the atlas of $M_6$ each leaf is homeomorphic
to a vector subspace of $\rr^6$. A {\it distribution} $D$ is a
choice at every point $x\in M_6$ of a 
subspace $D_x$ of the tangent space $T_x$ at that point.
A distribution is said to be {\it integrable} if there exists a foliation of $M_6$ such that, at every point $x$, $D_x$ is the tangent space to a leaf of the foliation. According to Frobenius' theorem, a distribution is
integrable if it is closed under Lie bracket. A less used, dual form
of this theorem can be given by looking at the orthogonal of $D$
with respect to the natural pairing between $T$ and $T^*$, 
$D^\perp \subset T^*$: namely, $D$ is integrable
if $D^\perp$ is closed under the action of $d$.

Consider now the distribution defined on $M_6$ by
the image of $P$, ${\rm im}(P)\subset T$. A classic result is that if $P$ is a Poisson tensor,  ${\rm im}(P)$ is an integrable distribution. We will
see later how to show this using pure spinors. 

Given a pure spinor $\Phi$, one can associate to it a matrix ${\cal J}$
acting on sections of $T\oplus T^*$, such that ${\cal J}^2=-1$. Concretely, in dimension 6 this is
a 12$\times$12 matrix that can be decomposed in 6$\times$6 blocks:
\begin{equation}\label{eq:blocks}
    {\cal J}=
    \left(\begin{array}{cc}
        I& P\\J&-I^t
    \end{array}\right)\ .
\end{equation}
(The lower--right block is not independent because of a certain
hermiticity property of ${\cal J}$ that follows automatically from
the non--degeneracy property of $\Phi$.) The way one associates such
a matrix to $\Phi$ is that the annihilator in $T\oplus T^*$ of $\Phi$
(which has dimension 6, by definition) is the $i$--eigenspace of
${\cal J}$; the annihilator of $\bar \Phi$ is the $-i$--eigenspace
of ${\cal J}$. For example, we saw in section \ref{sec:review} that a decomposable non--degenerate complex three--form $\Omega$ defines an almost complex structure $I$. Then one can see that
${\cal J}_I={{I \ \ \ 0}\choose {0\ -I^t}}$.

One can also show that if
\begin{equation}\label{eq:dPhi0}
    d \Phi=0
\end{equation}
then ${\cal J}$ satisfies a certain integrability condition. In
this case ${\cal J}$ is called a {\it generalized complex structure}.\footnote{Because of (\ref{eq:dPhi0}), one can
say that this generalized complex structure has a holomorphically
trivial canonical bundle, in analogy with the particular case
of an ordinary complex structure that we will consider shortly.
\cite{hitchin-gcy} calls this case ``generalized Calabi--Yau''.}
 We recognize
here the first condition in (\ref{eq:t-reform}). As we anticipated there, it follows that a RR solution is also a generalized complex
manifold.

As an example, we can consider an ordinary complex structure $I$.
We saw in section \ref{sec:review} that a non--degenerate complex
three--form $\Omega$ defines an almost complex structure $I$.
Then if one has (\ref{eq:dPhi0}), namely $d \Omega=0$, the
matrix ${\cal J}_I={{I \ \ \ 0}\choose {0\ -I^t}}$ defined above
should be a generalized complex structure. One can see that this implies that $I$ is a complex structure. It was indeed alredy known
well before the definition of generalized complex geometry that
$d \Omega=0 $ implies that the associated $I$ is a complex structure.

Going back to the general case, if
${\cal J}$ is generalized complex, it also follows that the upper--right
block $P$ in (\ref{eq:blocks}) is a Poisson tensor \cite{lmtz}.
In fact, we can see how the integrability of ${\rm im} (P)$ follows
from $d \Phi=0 $. Here is how. We know from (\ref{eq:exp}) the
general expression of a pure spinor of type $k$. It is easy to see
that the $\xi_i$ in that formula generate ${\rm ker}(P)$, which is
a subspace of $T^*$. Indeed, $\xi_i\wedge$ is in the annihilator of $\Phi$,
and hence $(0,\xi)\in T\oplus T^*$ must be an eigenvector of ${\cal J}$. Looking at (\ref{eq:blocks}), we see that this is the case
only if $P \xi_i=0 $, $\forall i=1,\ldots, k$. As we mentioned 
earlier, a dual form of Frobenius' theorem says that $D$ is integrable
if $D^\perp$ is closed under the action of $d$, where
the orthogonality is intended with respect to the natural pairing
between $T$ and $T^*$. In our case, 
$D={\rm im}(P)$, and $D^\perp= {\rm ker}(P)$.  
From $d \Phi=0$ it follows that
$d(\xi_1\wedge\ldots\wedge \xi_k)=0$, which means that $\{ \xi_i \}$
is closed under $d$; but $\{ \xi_i \}={\rm ker}(P)$. 
By the dual Frobenius theorem above, then, we have that ${\rm im}(P)$ is an integrable distribution, as we wanted to show.

With a pair of pure spinors, there are two Poisson
tensors $P_\pm$ that one might want to consider. In our case, 
$\Phi_\pm$ in
(\ref{eq:Bdielpuretilde}) have types 0 and 1 (in fact, we argued
in section \ref{sub:def} that they are the most general pair of
those types).  Hence $P_+$, the Poisson tensor associated to $\Phi_+$,
has no kernel. $P_-$, associated to $\Phi_-$, has kernel of dimension 1.

It is easy to recognize for example what $P_+$ is: since it is
invertible, we can define the two--form $P_+^{-1}$. When a
Poisson tensor is invertible, a consequence of (\ref{eq:Poisson})
is that its inverse two--form is closed. Hence $d(P_+^{-1})=0$.
Then it is easy to guess that $P_+^{-1}$ is nothing but the exponent
of $\Phi_+$. (We assumed in this section that all pure spinors
are closed; this condition implies that the upper--right block of (\ref{eq:blocks}) is Poisson. The generalized \ka\ condition (\ref{eq:int}) would also be sufficient.)

Actually, we can also associate $P_\pm$ to the pure spinors $B$--transformed in such a way as to have $B=0$; this means for us
the pair in (\ref{eq:dielpure}). Remember (from (\ref{eq:otimes}))
that in that case the pair can be written as $\Phi_\pm = \eta^1_+\otimes \eta^{2\,\dagger}_\pm$. Each of the two spinors
$\eta^{1,2}_+$ defines an almost complex structure $I_{1,2}$ (by looking at the gamma matrices that annihilate them, for example:
$(1+iI_{1,2})^n{}_m \gamma_n \eta^{1,2}_+=0$).
An expression for the Poisson tensors $P_\pm$ associated to
$\Phi_\pm$ is then \cite{gualtieri}\footnote{These Poisson
tensors were previously considered in \cite{lyakhovich-zabzine}.}
\begin{equation}
    P_\pm = (I_1 \pm I_2)g^{-1}\ .
\end{equation}
There is, however, a third Poisson tensor that one can consider
\cite{hitchin-instantons}:
\begin{equation}\label{eq:sigma}
    \sigma\equiv [I_1,I_2]g^{-1}
\end{equation}
and this is the one that will be important for us in what follows.
As noted in \cite{lindstrom-rocek-vonunge-zabzine}, since
$[I_1,I_2]= (I_1-I_2)(I_1+I_2)$, one has that $\ker [I_1,I_2]=
\ker (I_1+I_2) \oplus \ker (I_1-I_2) $. One can then decompose
the tangent space $T$ as
\begin{equation}\label{eq:decomp}
    \ker(I_1-I_2)\oplus \ker(I_1+I_2)\oplus {\rm im}(\sigma)
\end{equation}
and the last distribution is integrable, being the image
of a Poisson tensor.

\subsection{Sigma--model interpretation}
\label{sub:sigma}

The  decomposition (\ref{eq:decomp}) has a physical interpretation
that will be important for us. Remember that, if the two pure spinors
$\Phi_\pm$
are closed (as in (\ref{eq:N=2})), we are describing an $\nn=2$ vacuum.
Since there are no RR fields around, we can write down the sigma
model with this background as a target; this sigma model turns out
to have (2,2) supersymmetry on the worldsheet. This condition is
weaker than the one for an $\nn=2$ vacuum, as we saw around
equation (\ref{eq:int}).
Suppose this supersymmetric model can be written in terms of (2,2) superfields. There are three types of known superfields that include
scalars: {\it chiral} multiplets (the usual ones, that describe for example a sigma model with a \ka\ target space), {\it twisted chiral} multiplets, and {\it semichiral} multiplets (see for example \cite{gates-hull-rocek}). Suppose there are in our model respectively
$d_c$, $d_t$ and $d_s$ of each type. Then one can see that
\begin{equation}\label{eq:decomp-mult}
    2 d_c= \dim(\ker(I_1-I_2))\ , \qquad 2 d_t=\dim(\ker(I_1+I_2))\ , \qquad
    4d_s= \dim({\rm im}(\sigma))\ .
\end{equation}

In our case, since the pure spinors (\ref{eq:Bdielpuretilde}) have
types 0 and 1 respectively, we know already that $P_\pm$ have kernels
of dimensions 0 and 1.\footnote{If we call $z$ the one--form part of $\Phi_-$, which is then in the kernel of $I_1-I_2$, it also follows
that $(I_1 + I_2)z$ is in $\ker(I_1-I_2)$; since by assumption this
kernel is 1--dimensional, it follows that $I_1 z= I_2 z=i z$.\label{foot:z10}}
 Looking at (\ref{eq:decomp}), we see that
$\sigma$ must have an image of dimension 4. Looking at (\ref{eq:decomp-mult}), we see that this corresponds to having one
semichiral multiplet. Whether the remaining multiplet is a chiral
or twisted chiral multiplet is a matter of convention, and
we choose it to be chiral. In conclusion, (\ref{eq:Bdielpuretilde})
corresponds on the worldsheet to having one chiral and
one semichiral multiplet.

The paper \cite{lindstrom-rocek-vonunge-zabzine} gives the form
of the action in terms of (2,2) superfields for a model with an arbitrary number of chirals,
twisted chirals and semichirals. This gives a local construction
of generalized \ka\ manifolds. In section \ref{sec:gk} we will see how
to promote this to a full--blown $\nn=2$ supergravity solution.

\subsection{Restricting to four--dimensional leaves} 
\label{sub:leaf}

Before we do that, we need some more information on the Poisson
tensor $\sigma$ defined in (\ref{eq:sigma}).

We have learned that any Poisson tensor $P$ defines a foliation.
Since the tangent space to that foliation is given by the
distribution ${\rm im}(P)$, it follows that $P$ is invertible
on each leaf.

In the case of $\sigma$, we know that its leaves are
four--dimensional, and that they are given by $\{ w=w_0\}$ (remembering the discussion in section \ref{sub:branes}, especially (\ref{eq:d3})
and (\ref{eq:zw})).\footnote{Strictly speaking, the {\it generic}
leaves are four--dimensional; in the locus in which the one--form part of $\Phi_-$
vanishes, which is for us given by critical points of $w$,
the two complex structures coincide, and $\sigma$ vanishes.}.
Using local coordinates, one can define a restriction $\sigma_{(4)}$
of $\sigma$ to the leaf $\{ w=w_0\}$. In \cite{hitchin-instantons}, it
is shown that $\sigma$ is of type $(2,0)+(0,2)$; the $(2,0)$ part
with respect to $I_1$ ($I_2$) is holomorphic
with respect to $I_1$ ($I_2$). These properties remain true
for $\sigma_{(4)}$, which is also invertible. The two--form $\sigma_{(4)}^{-1}$ is holomorphic and of types $(2,0)+(0,2)$.
If we now take the $(2,0)$ part with respect to $I_1$, we have
a holomorphic $(2,0)$--form. A holomorphic version of the Darboux
theorem implies that there exist coordinates $q,p$ on $\{ w=w_0\}$
such that the $(2,0)$ part of $\sigma_{(4)}^{-1}$ is equal to
$dq \wedge dp$ (up to a constant). In particular, $q$ and $p$ are holomorphic coordinates
with respect to $I_1$. Similar coordinates exist for $I_2$, which
we call $Q,P$.

The preceding paragraph condenses some arguments in \cite{lindstrom-rocek-vonunge-zabzine}, to which we refer the reader
for more details; but the upshot for us is that one can define
two complex structures $I_{1,2}^{(4)}$ on the four--dimensional leaves, just by using the holomorphic coordinates above.

Now we can use that, in four dimensions \cite{hitchin-delpezzo},
\begin{equation}\label{eq:anticomm}
    \{I_1^{(4)},I_2^{(4)}\}=p 1_4\
\end{equation}
for some function $p$.
From this it also follows $[I_1^{(4)},I_2^{(4)}]^2=(p^2-4)1_4$.

Each of the complex structures $I_{1,2}^{(4)}$ on the leaf
define, together with the pullback of the metric $g_{(4)}$,
an SU(2) structure. The two--forms for these two SU(2) structures
satisfy the same relations as in (\ref{eq:su3i}), only now
with every form replaced by its four--dimensional counterpart.
Also, since $\sigma_{(4)}$ is invertible, we can write $g_{(4)}=\sigma_{(4)}^{-1}[I_1^{(4)},I_2^{(4)}]$.
We can now compute
\begin{equation}
    j_2^{(4)}=g_{(4)} I_2^{(4)}= \sigma_{(4)}^{-1}[I_1^{(4)},I_2^{(4)}]I_2^{(4)}=
    -\sigma_{(4)}^{-1}(2I_1^{(4)} +p I_2^{(4)}) \ ,
\end{equation}
and similarly for $j_1^{(4)}$; by using
then the four--dimensional analogues of (\ref{eq:su3i}) we can compute ${\rm Im }\omega_{(4)}$. One gets an
expression containing a symmetric part; setting it to zero one
obtains
\begin{equation}\label{eq:p-omi}
    \tan^2(2 \psi)=\frac{2+p}{2-p}\ , \qquad {\rm Im} \tilde\omega_{(4)}=
    2  \sigma_{(4)}^{-1}\ .
\end{equation}
Recall that the tilde means division by $\sin(4 \psi)$, as in (\ref{eq:tom-sp}).

(\ref{eq:p-omi}) now gives  
\begin{equation}\label{eq:omi4}
    {\rm Im} \tilde \omega_{(4)}= -\frac 12 (dq\wedge dp + d\bar q \wedge d \bar p)=-\frac 12(dQ\wedge dP + d\bar Q \wedge d \bar P)
\end{equation}
and
\begin{equation}
    i\,\tilde \omega^1_{(4)}= dq \wedge dp \ , \qquad
        i\,\tilde \omega^2_{(4)}= dQ \wedge dP \ .
\end{equation}
We will see in section \ref{sub:pot} how these equations
are extended to six dimensions.


\subsection{Noncommutativity} 
\label{sub:nc}

Poisson tensors are used in classical mechanics to define
Poisson brackets of functions on  phase space. A natural
question to ask is whether one can define naturally a noncommutative
product among those functions, or in other words whether it is
possible to quantize the Poisson bracket.

It has been argued for example in \cite{wijnholt} that this
noncommutativity should be related to the one defined by
the F--terms in the field theory duals, which we will review
in section \ref{sec:lm}. An immediate question
is why string theory should quantize the Poisson tensor
implicit in the geometry. One mechanism known in string theory
to produce noncommutativity is via a $B$--field
on the world--volume of a brane. In the case of a constant $B$--field
along a flat--space brane, the noncommutativity parameter is given by
\cite{seiberg-witten-nc,schomerus}
\begin{equation}\label{eq:theta}
    \theta= \frac12 [(g+B)^{-1}-(g-B)^{-1}]\equiv
    [(g+B)^{-1}]_{\rm A}\ .
\end{equation}

What we will show now is that
if one computes (perhaps naively) this tensor $\theta$ for the
D7 branes extended along a four--dimensional leaf $\{ w= w_0 \}$,
one obtains exactly the Poisson tensor $\sigma_{(4)}$ we were dealing with in section \ref{sub:leaf}. We find this very suggestive, and one might consider it as a generalization of the ``canonical coisotropic''
brane in \cite{kapustin-witten}\footnote{We thank E.~Witten for this remark.}; it also seems to give an alternative view of \cite{kapustin-top-nc}.
 On the cautionary side,
the role of $\theta$ as ``noncommutativity parameter'' is far from
clear in the non--flat case, and in any case
the computation below does not show why the noncommutativity should be transferred to
the field theory dual (a matter on which we will return in section
\ref{sec:lm}). Another problem is that a D7 wrapping the leaf
$\{ w=w_0\}$ appears to be supersymmetric for a $B'=B+ d \lambda$ different from the $B$ we are using in this paper (as we remarked in
section \ref{sub:branes}).

Be that as it may, pulling back (\ref{eq:B}) and using (\ref{eq:p-omi})
gives
\begin{equation}
    \label{eq:B4si}
    B_{(4)}=4 \sin^2(2 \psi)\sigma_{(4)}^{-1}\ .
\end{equation}
Recall again that, on $\{ w=w_0 \}$, $\sigma_{(4)}$ is invertible, and 
hence we can  rewrite (\ref{eq:sigma}) as $g_{(4)}= 
\sigma_{(4)}^{-1}[I^{(4)}_1,I^{(4)}_2]$. We
can then compute, with some manipulation, $\theta$ as defined
by (\ref{eq:theta}), with $g_{(4)}$ and $B_{(4)}$ as inputs:
\begin{equation}
    \label{eq:theta4}
    \theta=
    (2\sigma_{(4)}^{-1}(1+ I^{(4)}_1 I^{(4)}_2))^{-1}_{\rm A}=
    \frac1{8\sin^2(2 \psi)}((1+I^{(4)}_2 I_1^{(4)})\sigma_{(4)})_{\rm A}=
    \frac14 \sigma_{(4)}\ ,
\end{equation}
which is what we claimed.


\section{$\nn=2$ NS solutions from superspace}\label{sec:gk} 

We will use the worldsheet construction of generalized \ka\ structures
in \cite{lindstrom-rocek-vonunge-zabzine}, and find that a single
equation, (\ref{eq:det}), is enough to promote them to $\nn=2$ supergravity solutions.
We will do this in section \ref{sub:gcy} by computing the pure spinors
(\ref{eq:pureK}) associated to their
``generalized \ka\ potential'' (which we will review in section \ref{sub:lrvz}). After checking in section \ref{sub:checking} that
the pure spinors indeed satisfy the 
conditions for $(2,2)$ model, we will
impose the stronger condition for an $\nn=2$ supergravity vacuum
in section \ref{sub:gcy}.

\subsection{Generalized \ka\ manifolds from a potential} 
\label{sub:lrvz}

We will first review some more results we need from the paper \cite{lindstrom-rocek-vonunge-zabzine} (some part of it was
already reviewed in section \ref{sub:leaf}).

It is well--known that the off--shell supersymmetric action for a $(2,2)$ model without flux can be
written as an
integral over superspace of a single function real function $K$:
\begin{equation}\label{eq:K}
    S=\int d^2 \sigma d^2 \theta d^2 \bar\theta \, K\
\end{equation}
defined by $J=i\del\bar\del K$. The function $K$ depends on chiral
$(2,2)$ multiplets.

There exist however more general $(2,2)$ models, those whose target
space is a generalized \ka\ manifold (as briefly reviewed by us around
(\ref{eq:int})). For these more general target spaces, 
it has been known for some time \cite{gates-hull-rocek} that, to write the action in an off--shell
supersymmetric fashion, one also needs new multiplets called twisted
chiral and semichiral (as we mentioned already in section 
\ref{sec:geom}, see for example (\ref{eq:decomp-mult})). The action can still be written
as (\ref{eq:K}), but now $K$ is a function of all three types of multiplets, and not only of chirals only. The paper
\cite{lindstrom-rocek-vonunge-zabzine} found a geometrical interpretation of the ``generalized \ka\ potential''$K$, showing
in the process that it, and an off--shell action, exist locally for any
generalized \ka\ manifold and hence for any $(2,2)$ model.

Before we explain that interpretation, let us specialize our discussion
to the number and types of multiplets we need.
We reviewed in the previous section (see discussion below (\ref{eq:decomp-mult})) that a compatible pure spinor pair of types
0 and 1 (for which (\ref{eq:Bdielpuretilde}) is the most general
expression, as argued in section \ref{sec:dielectric}) corresponds
to having one semichiral and one ordinary chiral (2,2) multiplets.

There are two
different complex structures $I_{1,2}$ in a generalized \ka\ geometry,
as we saw in section \ref{sec:geom}. In section \ref{sub:leaf} we chose  holomorphic coordinates
for both $I_1$ and $I_2$. Since $\ker(I_1-I_2)=1$, or in other words,
there is one chiral multiplet, $I_1$ and $I_2$ share an eigenform
(see footnote \ref{foot:z10}); hence we took one of the holomorphic coordinates for $I_1$ and one of the holomorphic coordinates for $I_2$
to coincide.
We called it $w$, since, as we reviewed in section \ref{sub:branes},
it is the superpotential for a single D3 probe.
Following \cite{lindstrom-rocek-vonunge-zabzine}, we called $q,p$ the other two holomorphic coordinates for $I_1$; and $Q,P$ the
other two for $I_2$. Obviously these four are redundant,
and we will take $q$ and $P$ to be independent.

Now for the geometrical interpretation of $K$ in (\ref{eq:K}). On each leaf, the
transformation between $\omega^1_{(4)}$ and $\omega^2_{(4)}$
preserves the form ${\rm Im} \omega_{(4)}$. This form is closed
because of (\ref{eq:omi4}), and non--degenerate because (\ref{eq:omt})
is preserved by pull--back.
In other words, on each leaf the change of coordinates between $q,p$ and $Q,P$
is a canonical transformation with respect to ${\rm Im} \omega_{(4)}$. Now, $K$ in (\ref{eq:K}) is shown in \cite{lindstrom-rocek-vonunge-zabzine} to be the ``generating function'' of this canonical transformation, in the sense
that
\begin{equation}\label{eq:pQ}
    p=\del_q K \ ,\qquad Q=\del_P K\ ,
\end{equation}
just like in classical mechanics.

We will actually see in the next subsection that an alternative
definition exists: rather than being a function such that $J=i\del \bar \del K $ as in the \ka\ case, it is a function such that $\Phi_+=\exp[-\del\bar\del K]$, as we will see in (\ref{eq:pureK}).
We should notice that, although our focus is on $\nn=2$ supergravity
solutions, this particular result is valid locally for all generalized \ka\ manifolds, as we will explain after (\ref{eq:pure-lrvz+}).


\subsection{Pure spinors from a potential} 
\label{sub:pot}

We will now compute the pure spinors corresponding to the construction
of generalized \ka\ manifolds in \cite{lindstrom-rocek-vonunge-zabzine}
that we reviewed in section \ref{sub:lrvz}.

First some preliminary definition.
Given that there are two different complex structures, there are
two different Dolbeault operators $\del_1$, $\del_2$ that it 
would be natural to work with. In what
follows, somewhat surprisingly, it will be useful to consider the
``mixed'' Dolbeault operator
\begin{equation}
    \label{eq:del}
    \del\equiv dq\del_q+ dP\del_P+dw\del_w\ .
\end{equation}
which utilizes a holomorphic coordinate with respect to the complex
structure $I_1$ and one with respect to $I_2$.

Given this $\del$, we can compute the two--form $\del\bar\del K$. We will
also need the hermitian matrix of its coefficients in the coordinates $q,P,w$:
\begin{equation}\label{eq:KH}
    K_H=
    \left(\begin{array}{ccc}\vspace{.2cm}
        K_{q\bar q}& K_{q\bar P}&K_{q\bar w}\\
        K_{P\bar q}& K_{P\bar P}&K_{P\bar w}\\
        K_{w\bar q}& K_{w\bar P}&K_{w\bar w}
    \end{array}\right)\
\end{equation}
where for example $K_{q\bar q}=\del_q \del_{\bar q}K $.
It will also be useful to define its matrix of minors $R$. This is
the same as
\begin{equation}\label{eq:R}
    R=\det(K_H) K_H^{-1}\ .
\end{equation}
So for example
\begin{equation}
    R_{w\bar w}=K_{q\bar q}K_{P\bar P}-K_{\bar q P}K_{q \bar P} \ .
\end{equation}

Now for the computation of the pure spinors. First of all we can compute the trigonometric functions present
in the pure spinor Ansatz of the previous section.
This is done by taking the anticommutator $\{I_1,I_2\}$ (the
complex structures $I_{1,2}$ are explicitly given in \cite{lindstrom-rocek-vonunge-zabzine})
and restricting
it to the four semichiral directions. In those directions, it has
to be proportional to the identity \cite{hitchin-delpezzo}, as we
saw in (\ref{eq:anticomm}); from (\ref{eq:p-omi}) we find that the proportionality
factor $p=2(\sin^2(2\psi)-\cos^2(2\psi))$. This gives
\begin{equation}
    \label{eq:sincos}
    \sin^2(2 \psi)= \frac{K_{q\bar q}K_{P\bar P}-
    K_{\bar q P}K_{q \bar P}}{K_{qP}K_{\bar q \bar P}-K_{\bar q P}K_{q
    \bar P}}\ , \qquad
    \cos^2(2 \psi)=\frac{K_{ q P}K_{\bar q \bar
    P}-K_{q \bar q}K_{P \bar P}}{K_{qP}K_{\bar q \bar P}-
    K_{\bar q P}K_{q \bar P}}\ .
\end{equation}

Now a few remarks about the metric, which is given
by \cite{lindstrom-rocek-vonunge-zabzine} in their (3.33), which we can easily specialize to the case with one semichiral and one chiral multiplet. We saw in section \ref{sec:geom}
that the Poisson tensor $\sigma$ defines a foliation whose leaves
are $\{ w=w_0\}$.
If we stay away from the
critical loci of $w$, where the leaves change dimension, this foliation
is just a fibration, and we can write the metric in a way adapted
to it:
\begin{equation}\label{eq:fibr}
    g= g_{ij}dy^i dy^j +2g_{ia}dy^i dx^a +g_{ab}dx^a dx^b=
    g_{ij}(dy^i +A^i{}_a dx^a)\cdot
    (dy^j +A^j{}_b dx^b)+ z\cdot \bar z\ .
\end{equation}
Here  $\cdot$ denotes
the symmetric tensor product, and we have denoted the coordinates $q, P, \bar q, \bar P$  collectively by $y_i$, and $w, \bar w$ by $x_a$. The right hand side of (\ref{eq:fibr}) defines $z$. Any metric can
be rewritten in the form (\ref{eq:fibr}); the presence of the foliation guarantees that this
is globally well--defined (away from the critical loci of $w$),
although this is not going to be too important for us, since we
are eventually going to apply these metric to $\rr^6$.
From the explicit form of the metric given in \cite{lindstrom-rocek-vonunge-zabzine},
one can compute $A$ and $z\cdot\bar z$ in (\ref{eq:fibr}):
\begin{equation}\label{eq:A}
    A^q=-\frac{R_{w\bar q}}{R_{w\bar w}}
    dw \ , \qquad
    A^P=-\frac{R_{w\bar P}}{R_{w\bar w}}
    dw\ ,
\end{equation}
along with $A^{\bar q}=\overline{A^q}$ and
$A^{\bar P}=\overline{A^P}$, and
\begin{equation}\label{eq:z}
    z\cdot\bar z =
    4\frac{\det(K_H)}{R_{w\bar w}}
    dw\cdot d\bar w\equiv \rho^2 dw\cdot d\bar w .
\end{equation}
$g_{ij}$ is not particularly interesting and we do not need
its explicit expression.
($A$ can also be extracted from $\{I_1,I_2\}$.)

We now want to write down the pure spinors for the generalized \ka\
metric in \cite{lindstrom-rocek-vonunge-zabzine}, using the
complex structures $I_{1,2}$ that they give explicitly in their formulas (6.72,6.74). To do this, one
can proceed in several ways. One can for example compute
$J_1=g I_1$ and $J_2=g I_2$, and then use (\ref{eq:Jj}) and sums
and differences of (\ref{eq:su3i}) to compute $\omega_r$, $\omega_i$ and $j$, and then $\omega^1$ and $\omega^2$. Without giving more
details, we will describe here the result. Define the following
``push--forward'' way to extend forms from the four dimensions spanned by $q,P$ to
the whole six--dimensional manifold:
\begin{equation}\label{eq:iota}
    \iota_* (d_4 q) \equiv Dq\equiv dq+ A^q\ , \qquad
    \iota_* (d_4 P) \equiv DP\equiv dP+ A^P\ ,
\end{equation}
where $d_4$ is the exterior differential along the leaves $\{w=w_0 \}$.
One has, then,
\begin{equation}
    \label{eq:om6}
    \frac i{\sin(4 \psi)}\omega^1=\iota_* (d_4 q\wedge d_4 p)\ ,\qquad
    \frac i{\sin(4 \psi)}\omega^2=\iota_* (d_4 Q\wedge d_4 P)\ .
\end{equation}
To evaluate these expressions concretely, one should first express
$Q=\del_P K$ and $p=\del_q K$, and then use (\ref{eq:iota}), which
results in
\begin{equation}\label{eq:om6con}
    \begin{array}{cc}\vspace{.2cm}
        \frac i{\sin(4 \psi)}\omega^1=8 Dq\wedge(K_{q\bar q}D\bar q
        +K_{qP}DP+ K_{q\bar P}D\bar P)\ , \\
        \frac i{\sin(4 \psi)}\omega^2=8(K_{Pq}Dq+
        K_{P\bar q}D\bar q+ K_{P\bar P}D\bar P) \wedge DP\ ;
    \end{array}
\end{equation}
one can see that with this definition
$\tilde\omega^a=\omega^a/\sin(4 \psi)$ satisfy the conditions (\ref{eq:omt}) and (\ref{eq:psi}). One can also see that (\ref{eq:om6con})
imply, for the $\Omega^{1,2}$ defined in (\ref{eq:Jj}),
\begin{equation}\label{eq:Om6}
    \begin{array}{cc}\vspace{.2cm}
        \frac i{2\cos^2(2 \psi)}\Omega^1=
        8 dw\wedge dq\wedge dp \ , \\
        \frac i{2\cos^2(2 \psi)}\Omega^2=
        8 dw\wedge dQ\wedge dP \ ,
    \end{array}
\end{equation}
after some simplification recalling that the coefficients in $A^q$, $A^P$ come from the inverse matrix of $K_H$ (see
(\ref{eq:R})). $\Omega^{1,2}$ are the $(3,0)$--forms with respect to the two complex structures $I_{1,2}$; since the latter are integrable,
$\Omega^{1,2}$ had to be indeed conformally closed. Even more precisely, we see from (\ref{eq:dil2}) that $d(e^{-2 \phi} \Omega^{1,2})=0$. This agrees with the computation in \cite{gauntlett-martelli-waldram} in the case of ${\cal N}=1$
backgrounds, applied to $\eta^{1,2}$ separately. Let us
stress again that in expressions such as (\ref{eq:Om6}),
$p$ and $Q$ are not independent variables, and are given by (\ref{eq:pQ}). So more explicitly
\begin{equation}
dw\wedge dq \wedge
dp= dw \wedge dq\wedge
(K_{q\bar q}d\bar q + K_{qP}dP + K_{q\bar P} d\bar P+ K_{q\bar w}d\bar w)   \ .
\end{equation}

Another combination of $\omega^{1,2}$ that can be simplified
considerably is their difference:
\begin{equation}\label{eq:diffom}
    \frac i{s_{4\psi}}(\omega^1-\omega^2)+
        \frac12 z\bar z=8\del\bar\del K \ .
\end{equation}

Putting now together (\ref{eq:Bdielpuretilde}),
(\ref{eq:Om6}) and (\ref{eq:diffom}) we get
\begin{eqnarray}
\label{eq:pure-lrvz+}   &&
    \Phi_+=\exp[8\,\del\bar\del K] \ ,\\
\label{eq:pure-lrvz-}   &&
    \Phi_-=\rho \tan(2 \psi) dw\wedge\exp\Big[
        8\, dq \wedge d(\del_q K) +
        8\, d(\del_P K) \wedge dP \Big]\ .
\end{eqnarray}
where we have left the prefactor in $\Phi_-$ unspecified. This
prefactor will be our focus in section \ref{sub:gcy}. For now, however,
notice that the pair we have just obtained is applicable locally
to any generalized \ka\ manifold, since we have
not yet derived nor imposed the extra conditions for it to be an
$\nn=2$ supergravity vacuum. In particular, this gives a
possible definition of $K$ which is formally identical to the
more usual \ka  \ definition.

A curious parallel to this situation
exists for topological theories. If one defines the A model on a \ka\ manifold, it is well--known \cite{witten-mirror-topological}
that the action can be written as $S=\{Q,V \}+ W$, with $W= \int J $
a ``topological'' action. The A model on a generalized \ka\ manifold
has, as topological action, precisely the two--form in the exponent of an even pure spinor \cite{zucchini-bihermitian-topological,chuang}.

The use of a potential to describe pure spinors also advocated in \cite{minasian-petrini-zaffaroni} for RR solutions; the reason 
a potential can still play a role in the RR case 
is that the real part of $\Phi_+$
is still (conformally) closed (see the second equation in (\ref{eq:t-reform})).

\subsection{Checking the generalized \ka\ condition}\label{sub:checking} 

Since the pure spinors (\ref{eq:pure-lrvz+}), (\ref{eq:pure-lrvz-})
have been derived from \cite{lindstrom-rocek-vonunge-zabzine},
they should define a generalized \ka\ structure by construction.
We will now check that they indeed
satisfy the conditions (\ref{eq:int}).

A pure spinor of type 0 can be written as $\Phi_+=e^\alpha$ for some two--form $\alpha$.
It is easy to see, then, that (\ref{eq:int}) can only be satisfied if
$d \alpha=0$. Indeed, (\ref{eq:int}) says that we should have
\begin{equation}\label{eq:int0}
d e^\alpha=(\iota_v + \xi\wedge)e^\alpha
\end{equation}
for some $v$ and $\xi$. The right hand side of (\ref{eq:int0}) can be rewritten as $(\iota_v \alpha+ \xi)\wedge e^\alpha$. Then the
zero--form part of (\ref{eq:int0}) says that $0=\iota_v \alpha+ \xi$, hence $d \alpha=0$.
In other words, the exponent in $\Phi_+$ in (\ref{eq:pure-lrvz+}) has to be closed. Checking this is immediate because $\del\bar\del K=
d(\del-\bar\del)K$.

We now come to $\Phi_-$. Again we take a step back and we ask ourselves what (\ref{eq:int}) says about a general pure spinor of type 1, namely  $\Phi_1\wedge e^\beta$ with $\Phi_1$ a one--form
and $\beta$ a two--form. Proceeding in a way similar to the type 0 case, one can reduce (\ref{eq:int}) to the conditions
\begin{equation}\label{eq:int1}
\Phi_1\wedge d \sigma=0 \ , \qquad d(f \Phi_1)=0 \ \ \ {\rm for\ some } \ f\ .
\end{equation}
For (\ref{eq:pure-lrvz-}), the first condition in (\ref{eq:int1})
is satisfied because the exponent is closed by itself. The second
condition is satisfied by taking $f=(\rho \tan(2 \psi))^{-1}$.


\subsection{Imposing the generalized \cy\ condition}\label{sub:gcy} 

We have explained in section \ref{sec:review} how the condition for an $\nn=2$ supergravity vacuum (\ref{eq:N=2}) is in general
stronger than the condition for $(2,2)$ worldsheet supersymmetry (namely, that the target space should be  a generalized \ka\ manifold). In the previous subsection we checked the generalized \ka\ condition, which is equivalent to worldsheet $(2,2)$ supersymmetry. We 
now want to see what remains
to be imposed for ${\cal N}=2$ supersymmetry in the target space,
which is condition (\ref{eq:N=2}).

We have noted in the previous subsection that actually this is already satisfied for $\Phi_+$ in (\ref{eq:pure-lrvz+}).
As for $\Phi_-$ in (\ref{eq:pure-lrvz-}),
we noted in the previous subsection that the exponent is already
closed; the one--form in front, however, is only conformally closed
(that is, up to a function, see (\ref{eq:int1})). This
is good enough for the generalized \ka\ condition to be true, but
not quite for (\ref{eq:N=2}). So we have to impose that
 $\rho\tan(2 \psi)$ in (\ref{eq:pure-lrvz-}) be a constant; remembering the definition of $\rho$ in (\ref{eq:z}) and the expressions
for the trigonometric functions in (\ref{eq:sincos}), we get
\begin{equation}\label{eq:det}
    \begin{picture}(50,20)(-25,0)
        \put(-20,-20){\line(1,0){180}}
        \put(-20,25){\line(1,0){180}}
        \put(-20,-20){\line(0,1){45}}
        \put(160,-20){\line(0,1){45}}
    \end{picture}\hspace{-1cm}\frac{\det(K_H)}{K_{qP}K_{\bar q \bar P}-K_{q\bar q}K_{P\bar P}}={\rm const}\ . \vspace{.3cm}
\end{equation}
Recall that $K_H$ is a matrix given in (\ref{eq:KH}). This equation is a generalization of the usual Monge-Amp\`ere equation and
reproduces the expression derived from worldsheet techniques in \cite[Eq.(18)]{grisaru-massar-sevrin-troost}\footnote{A generalization of the Monge--Amp\`ere equation was also derived in \cite{rocek-modified-torsion}, which considers the case with 
chiral and twisted chiral multiplets.}.

The situation is very similar to the usual \ka\ case. There,
one can easily define a \ka\ metric on $\cc^3$ from a real function $K$, via $J=i\del\bar\del K$ and trivial complex structure.
One can define two pure spinors
\begin{equation}\label{eq:ka}
\Phi_+=e^{iJ}\ , \qquad \Phi_-=\det(\del\bar\del K) \, dz^1\wedge dz^2
\wedge dz^3\ ,
\end{equation}
which satify the algebraic constraints for two compatible pure spinors
(in particular they have equal norm). The conditions (\ref{eq:int})
are indeed satisfied, as should be the case since a \ka\ metric is
also generalized \ka; to obtain an $\nn=2$ supergravity solution
one also needs to impose that both pure spinors in (\ref{eq:ka})
are closed. This imposes that $\det(\del\bar\del K)={\rm const}$,
which is the \ka\ analogue of (\ref{eq:det}).

\bigskip\bigskip

To summarize: for any function $K(q,P,w,\bar q, \bar P, \bar w)$
that satisfies (\ref{eq:det}), the pure spinors
\begin{equation}\label{eq:pureK}
\fbox{$\begin{array}{l}\vspace{.3cm}
    \Phi_+=\exp[-8\del\bar\del K] \ ,\\
    \Phi_-= dw\wedge\exp\Big[
        -8 dq \wedge d(\del_q K) -
        8 d(\del_P K) \wedge dP \Big]
        \end{array}
        $}
\end{equation}
are compatible and satisfy (\ref{eq:N=2}). Adding some technical
requirements, namely that the metric should be actually positive
definite and nonsingular, and that the volume form defined by the pure spinors should
have no zeros (in our case, this requires that the function $\det(\del\bar \del K)$ have no zeros), (\ref{eq:pureK}) define
an ${\cal N}=2$ solution in type II supergravity.

At first sight, it looks like (\ref{eq:pureK}) gives now a very easy
way of producing solutions: one could think for example that any
set of coordinates $q, P, w, \bar q, \bar P, \bar w$ and any $K$
quadratic in those coordinates will do the job, since a quadratic $K$
will surely satisfy (\ref{eq:det}).\footnote{Another Ansatz one
could try to solve (\ref{eq:det}) is to take $K=f(w,\bar w)+ K_0(q,P,\bar q, \bar P)$. In this case, the determinant in the numerator of (\ref{eq:det}) factorizes, and one recovers \cite[Eq.(5.1)]{bogaerts-sevrin-vanderloo-vangils}, that describes
a hyper--K\"ahler manifold. We thank M.~Zabzine for this comment.}
In fact, however, such solutions
will correspond to flat metrics. To give a nontrivial example to
this construction, we will now resort to a known solution.



\section{A special case: Lunin--Maldacena NS solution} 
\label{sec:lm}

In this section we show that a certain solution given in \cite[Page 23]{lunin-maldacena} is described by the pure spinors in (\ref{eq:pureK}) for an appropriate choice of coordinates and $K$.

The solution is purely NS (we call it LM--NS solution from now on) and is very similar in form to another solution in \cite{lunin-maldacena}; this one has RR fields too, and we will call it simply LM solution.
The LM solution is the gravity dual to the so--called $\beta$--deformation
of $\nn=4$ SYM, one of the Leigh--Strassler theories \cite{leigh-strassler}.
We hence start with a quick review of those theories. We then
will show how one can derive the LM--NS solution in the pure
spinor formalism. We finally show how it is a particular case
of (\ref{eq:pureK}).

\subsection{Leigh--Strassler theories} 
\label{sub:ls}

 $\nn=4$ super Yang--Mills is a conformal theory and as such it is interesting to study the space of its
 exactly marginal deformations. The arguments of Leigh and Strassler \cite{leigh-strassler} rely on preserving a discrete ${\mathbb Z}_3$ symmetry which permutes the three $\nn=1$ chiral superfields $X_i$ and as a result there are exactly two candidate deformations:
 \begin{equation}\label{eq:wls}
    W={\rm Tr}\Big([X_1,X_2]X_3 + \gamma \{ X_1, X_2\}X_3
    + \gamma'\sum_{i=1}^3 X_i^3\Big)\ .
\end{equation}
 The LM solution is the gravity dual to the family of theories with $\gamma'=0$ and is sometimes also referred to as the {\it beta deformation}.

One of the original motivations for this paper was to find the gravity
dual for the most general theory with $\gamma'\neq 0$.  Some approximate results have already been
found, for example \cite{grana-polchinski,aharony-kol-yankielowicz, Kulaxizi:2006zc} in perturbation theory; these papers however also serve as an illustration of how complicated such a task can get at higher orders.

The strategy we want to promote in this paper is that one might make
progress by considering an ``auxiliary'' purely NS solution. The idea
is that the gravity dual will then arise by placing D3 branes on this NS solution, much like the relation between flat space and AdS$_5\times S^5$. Thanks to the work of Lunin and Maldacena \cite{lunin-maldacena} we know this is the case for $\gamma$ real and $\gamma'=0$. (The solution for $\gamma$ imaginary is then obtained by S--duality and as such is obviously a RR background.) As described below, the reformulation of this problem in terms of generalized complex geometry suggests that there should be in addition an NS solution for $\gamma'\neq 0$. 

We believe that some features
of the full RR gravity dual can already be captured by the NS solution. For example,
we saw in section \ref{sub:branes} that the moduli space of D3 branes
is now partially lifted: a single, spacetime--filling D3 brane
can only be placed at a critical point of the function we have called $w$.
If we go back to (\ref{eq:wls}), we see a similar feature. Unlike
the usual $\nn=4$ theory, this superpotential is non--zero even
when we set the rank of the $X_i$ to 1. $W$ becomes then a function
$w$
on $\cc^3$, and the supersymmetric vacua become its critical points.
By using this $w$ in
the construction (\ref{eq:pureK}), one would find (if one were able
to solve (\ref{eq:det})) a geometry such that a D3 brane probe would
have exactly $w$ as an effective potential.\footnote{Another way of finding a metric with this feature has been pointed out in \cite{maldacena-sheikhjabbari-vanraamsdonk}: it should be obtainable
by dualities from their equation (20). We thank J.~Maldacena for
pointing this out.}
 Hence, at least an NS precursor of the full gravity dual should be in the class (\ref{eq:pureK}). In section \ref{sub:lm_pot} we will see explicitly that this is indeed the case for the only known gravity dual.

Probe D3 branes should experience a superpotential $w$ in the full
RR solution too \cite{Corrado:2004bz}. Recall that a warped product $AdS_5\times M_5$ can also be considered as a warped product $R_{1,3}\times M_6$, with $M_6$ the cone over $M_5$. It was shown in \cite{gauntlett-martelli-sparks-waldram-ads5-IIB} that for any AdS$_5$ solution of IIB supergravity, $M_6$ must be either Calabi--Yau or have an ${\rm SU}(2)$ structure. Since by definition the gravity duals of Leigh--Strassler deformations have an AdS$_5$ factor and are not Calabi--Yau, we see that they must be ${\rm SU}(2)$ structure solutions. This means
that there are two complex structures $I_{1,2}$. As we stressed in section \ref{sec:geom}, this is the case when the pure spinor
$\Phi_-$ is of type 1, namely it has a non--zero one--form $dw$;
in that case, $w$ is then the superpotential for D3 brane probes
that we claimed to be a common feature to the NS and RR solutions.

Another possible check of this idea comes to mind \cite{wijnholt} after noticing that the F--term equations for (\ref{eq:wlm}) look like
\begin{equation}\label{eq:F}
\begin{array}{c}\vspace{.2cm}
    {}[X_1,X_2]=\gamma\{ X_1,X_2\}+3 \gamma' X_3^2\ , \\\vspace{.2cm}
    {}[X_2,X_3]=\gamma\{ X_2,X_3\}+3 \gamma' X_1^2\ , \\
    {}[X_3,X_1]=\gamma\{ X_3,X_1\}+3 \gamma' X_2^2\ .
\end{array}
\end{equation}
If one interprets this as a noncommutativity on $\cc^3$, one might
try to relate it to the computation in section \ref{sub:nc}, although
see the various caveats there.


\subsection{Moduli and Poisson bivectors} 
\label{sub:moduli}

We just proposed that the class of solutions considered in
section \ref{sec:gk} should be related to gravity duals
of Leigh--Strassler theories. In this subsection we supplement these arguments, based on observations
about the number of moduli in the field theory and in geometry.
As a by--product, we will also see a possible generalization
to more general conformal field theories.

One of the achievements of generalized complex geometry \cite{gualtieri} was the geometrical interpretation of Witten's extended moduli space \cite{witten-mirror-topological}. If one
starts from a complex manifold $M_6$, its infinitesimal
complex deformations are given by $H^1(M_6,T_{1,0})$. As we
saw in section \ref{sec:geom}, in the discussion after
(\ref{eq:dPhi0}), to a complex
structure $I$ one can associate a generalized complex structure
${\cal J}_I$; the associated pure spinor (which exists if $c_1=0$)
is the holomorphic three--form for $I$. It follows that, quite
reasonably, a complex manifold can be considered as a generalized
complex manifold. The infinitesimal deformations of ${\cal J}_I$
are given by
\begin{equation}\label{GCGModuli}
\oplus_{p+q=2}H^p(M_6,\Lambda^q T_{1,0})\ ;
\end{equation}
the obstructions to these moduli live in
\begin{equation}\label{GCGObstr}
\oplus_{p+q=3}H^p(M_6,\Lambda^q T_{1,0})\ .
\end{equation}
(\ref{GCGModuli}) are deformations of a complex
structure as a generalized complex structure; indeed, ordinary
complex structure deformations, that live in $H^1(M_6,T_{1,0})$,
are a subset of (\ref{GCGModuli}). The other two summands, $H^0(M_6,\Lambda^2 T_{1,0})$ and $H^2(M_6,{\cal O})$, vanish if $M_6$ is \cy; in that case, the only deformations are the more familiar complex structure deformations in $H^1(M_6,T_{1,0})$.

However, an interesting observation was made in \cite{wijnholt} by considering non--normalizable modes on $M_6$.
Consider a Calabi--Yau threefold $M_6$ which is a cone over a {\it regular} Sasaki--Einstein manifold $M_5$. This five dimensional manifold $M_5$ is by definition a ${\rm U}(1)$ bundle over a K\"ahler--Einstein space $B_4$. The observation in \cite{wijnholt} was that one could consider elements $\beta^{ij}\in H^0(B_4,\Lambda^2 T_{1,0})$ and then holomorphically extend these over the entire Calabi--Yau cone $M_6$ to obtain a non-commutative deformation.
Physically this is entirely reasonable since according to the AdS/CFT dictionary, non--normalizable modes in the bulk are dual to superpotential deformations (normalizable modes are dual to vev's). These bivector deformations should be dual to the exactly marginal deformations
of the field theory on a stack of D3 branes at the tip of this cone.
It has been checked \cite{wijnholt} in a few examples that the number of bivector deformations matches the total number of exactly marginal deformations, although one needs to restrict to {\it regular} Sasaki-Einstein spaces. The paper \cite{bergman} gives a general
reason for the match, and defines a map from bivectors to field theory
deformations. On the conifold, the gravity dual of one particular deformation with ${\rm SU}(2)$ global symmetry was found in \cite{Halmagyi:2005pn}, and in \cite{benvenuti-hanany} the space of exactly marginal deformations for the conifold was written down in the field theory following ideas in \cite{Kol:2002zt}.

In the case of AdS$_5\times S^5$, the
bivector deformations are the ones discussed by Leigh and Strassler.
For more general field theories, we can observe that a bivector $\beta^{ij}$ deforms the pure spinor $\Omega$ of a \cy\ as
\begin{equation}\label{eq:bivdef}
    \delta\Phi_-= \beta^{ij}\iota_{\del_i}\wedge\iota_{\del_i}\,
    \Omega\ ,
\end{equation}
at first order.
 This means that the action of $\beta^{ij}$ generates a one--form in
$\Phi_-$. Since for a vacuum this has to be closed (see (\ref{eq:t-reform})), it can be written locally as $\beta^{ij}\iota_{\del_i}\wedge\iota_{\del_j}\Omega=dw$.
As we argued in \ref{sub:branes} and again in \ref{sub:ls},
this means that D3 brane probes feel an effective superpotential $w$.

This would seem to confirm further the intuition we promoted in
section \ref{sub:ls}, namely that the gravity duals to Leigh--Strassler
theories should be obtained by placing D3 branes on a NS solution
of the type considered in section \ref{sec:gk}. In fact, it also
seems to indicate that such solutions should be relevant to deformations of CFT's other than $\nn=4$.

There are several obstacles, however, to using the observation
in \cite{wijnholt} directly. The first concerns obstructions, the
second concerns introducing a second pure spinor $\Phi_+$ in the story.
Although the two are connected, at first we will consider them
separately.

If one was ultimately concerned with the deformation of just $B_4$, then one would need only to calculate the obstruction group (\ref{GCGObstr})
to determine whether the deformation can be made for finite values of the parameter. This is indeed done in \cite{gualtieri} for $B_4={\mathbb CP}^2$. However, we are concerned not just with the generalized complex structure on $B_4$ but with the one on $M_6$.
The latter also has a holomorphically trivial canonical bundle, in that
there exists a closed pure spinor for it (in the terminology of
\cite{hitchin-gcy}, it is a generalized \cy\ structure).
We may then appeal to a theorem in \cite{hitchin-gcy} which tells us the moduli of a generalized Calabi-Yau manifold are integrable.

This issue is complicated, however, by the entrance in the scene of a second pure spinor $\Phi_+$. Remember from section \ref{sec:review}
that this is needed both for NS vacua (\ref{eq:N=2}) and for RR
vacua (\ref{eq:t-reform}). One has to find a way to promote the
bivector deformations (\ref{eq:bivdef}) to deformations of a compatible
pure pair $\Phi_\pm$.

One way to deform a compatible pair $\Phi^0_\pm$ is to act on it with
the same element $O \in {\rm O}(6,6)$; since such an element keeps the
internal product $(\cdot ,\cdot )$ in (\ref{eq:defMukai})
invariant, conditions 1.~and 2.~in the definition of compatibility in section \ref{sec:review} are
kept satisfied (condition 3.~is an open condition, so it is not
affected by small enough deformations). Next, since we are interested in deformations of $\Phi_-$, we might further restrict
deformations so that $\Phi_+$ is left unchanged. Remarkably, we will see in section \ref{sub:td},
and in particular in (\ref{eq:lm+}), that this is exactly what
happens for the LM--NS solution. At first order,
one can show  \cite[Section 6.5.1]{gualtieri} that this leads to
\begin{equation}\label{eq:gKdef}
\delta\Phi_{\pm}= {\rm Re} [\beta^{ij}
(\iota_{\del_i}+\frac12 J_{i\bar k} d\bar z^{\bar k})\wedge
(\iota_{\del_j}-\frac12 J_{j\bar l} d\bar z^{\bar l})] \Phi_{\pm}\equiv
{\rm Re} (\epsilon) \cdot\, \Phi^0_\pm \ ;
\end{equation}
the operator ${\rm Re} (\epsilon)$ is indeed in the Lie algebra ${\rm o}(6,6)$, and one can see that ${\rm Re} (\epsilon)\, \cdot\Phi_+=0$. The operator ${\rm Re} (\epsilon)$
is a linear combination of a bivector, of a tensor with one index
up and one down\footnote{This part is present only if $\beta^{ij}$ has a symmetric part too.}, and of a two--form. At first order,
the deformation (\ref{eq:gKdef}) yields a generalized \ka\ structure
(defined by (\ref{eq:int})),
if one chooses these three components to be respectively in $H^0(M_6,\Lambda^2 T_{1,0})$,
$H^1(M_6, T_{1,0} )$ and $H^2(M_6,{\cal O})$.

If one now tries to extend (\ref{eq:gKdef}) to all orders,
the most natural possibility is just to write
\begin{equation}\label{eq:fingKdef}
\Phi_{\pm}= \exp[{\rm Re} (\beta^{ij}
(\iota_{\del_i}+\frac12 J_{i\bar k} d\bar z^{\bar k})\wedge
(\iota_{\del_j}-\frac12 J_{j\bar l} d\bar z^{\bar l}))]\Phi^0_{\pm}
 \ .
\end{equation}
Since the exponential is now in ${\rm O}(6,6)$, these pure spinors
are now a compatible pair, as we noted above.
On the other hand, for simple $\beta^{ij}$ it can happen
that $\epsilon$ in (\ref{eq:gKdef}) squares to zero, even if
its real part ${\rm Re} (\epsilon)$ (which is in ${\rm o}(6,6)$)
does not. In this case one has another option. One can check that
the first order deformation (\ref{eq:gKdef}) is also
\begin{equation}
    \delta \Phi_{\pm}= \epsilon\cdot \Phi^0_{\pm}\ ;
\end{equation}
in other words, one can drop the ${\rm Re} $ in (\ref{eq:gKdef}).
One can then simply define $\Phi_\pm= e^\epsilon \Phi^0_\pm=
(1+\epsilon)\Phi^0_\pm $. For this tentative pair, condition 1.~for
compatibility is
still satisfied, because it only needs to be checked at first order,
where it is still true (because it was true for (\ref{eq:gKdef}),
in which case we can use that ${\rm Re} (\epsilon)\in {\rm o}(6,6)$).
Condition 2.~is not guaranteed this time; but one can simply
rescale each $\Phi_\pm$ by their new norm. Summarizing,
\begin{equation}\label{eq:altfingK}
      \Phi_\pm=
    \frac{(1+\epsilon\cdot)\Phi^0_\pm}{||(1+\epsilon\cdot)\Phi^0_\pm||}\qquad \qquad({\rm if}\  \epsilon^2=0)\ .
\end{equation}

The finite--order deformation (\ref{eq:fingKdef}) applies more generally. It also avoids the step of dividing by the norm as in
(\ref{eq:altfingK}),
so it is more appropriate for looking for supergravity vacua,  namely solutions to (\ref{eq:N=2}).
On the other hand, the second method, (\ref{eq:altfingK}), can be applied to generalized \ka\ manifolds, namely solutions to (\ref{eq:int}). In that case, a rescaling  $\Phi_\pm\to e^f \Phi_\pm$
can be reabsorbed by changing $\xi_\pm\to \xi_\pm - d\log(f)$ in
(\ref{eq:int}). (In fact, for
this very reason, for generalized \ka\ manifolds part 2.~of
the compatibility condition can be omitted.)

Both these ways of making
 (\ref{eq:gKdef}) finite were utilized in \cite{lin-tolman, bursztyn-cavalcanti-gualtieri} to produce examples of
generalized \ka\ structures for their general theory of generalized K\"ahler reduction. In this application, it is crucial that  $\Phi_+$ is invariant under (\ref{eq:gKdef}); morally, one can still use the two--form exponent of $\Phi_+$ for the symplectic quotient.
This reduction procedure should be inherent to the gravity duals of the Leigh--Strassler deformations, because it would give a way to count
the deformations more directly on the base $B_4$, as in \cite{wijnholt}.
The deformations considered on $\cc^3$ in \cite{lin-tolman,bursztyn-cavalcanti-gualtieri}, however,
satisfy (\ref{eq:int}) but {\it do not} satisfy (\ref{eq:N=2}), and hence do not give rise to $\nn=2$ vacua. The pure spinors they consider either fail to satisfy condition 2.~for compatibility (as given in section \ref{sec:review}), or, if they are divided by their norms as in (\ref{eq:altfingK}), are not closed. In section \ref{sub:lm_pot}, we will present the LM--NS solution as an example of (\ref{eq:gKdef}) which does, on the contrary, satisfy (\ref{eq:N=2}).

\subsection{T--duality and bivector action} 
\label{sub:td}

From now on we will consider the case of $\gamma'=0$ in eq. (\ref{eq:wls}).
The method used in \cite{lunin-maldacena} to generate their solutions
was to act with an element of the symmetry group of supergravity
${\rm O}(6,6)$ that does not belong to the stabilizer (isomorphic to ${\rm O}(6)\times {\rm O}(6)$) of the initial solution.

The action of T--duality on pure spinors was considered in \cite{minasian-petrini-zaffaroni}
by using ordinary spinors.
Here we want to present an alternative way of computing that action,
which is more in line with other geometrical ideas present in this paper.

The method is similar to the one explained in detail in \cite[Section 6.1]{gmpt3}. One exploits the fact that T--duality acts on $T\oplus T^*$. One first
computes the annihilator of the pure
spinor $\Phi$ we want to transform, ${\rm Ann}(\Phi)\subset T\oplus T^*$; then one acts on this annihilator with the T--duality element $O \in {\rm O}(6,6)$;
then one finds $\tilde \Phi$ such that ${\rm Ann}(\tilde \Phi)=O({\rm Ann}(\Phi))$.

The result of this procedure has a subtle part, the mixing of
the metric and $B$--field, and an easy part, the actual action on the pure spinor. The subtle part is that in general the manifold $M_6$ is changed into some new manifold $\tilde M_6$. This is source of much of the agony
in \cite[Section 6.1]{gmpt3}, where care is needed because the connection on the $S^1$--fibration on the original $M_6$ is exchanged
by T--duality with some component of the $B$--field.

Fortunately,
this kind of subtlety is not relevant for us. Indeed, let us consider
a general $T^3$--fibration with coordinates $r^i$ on the base and
$\phi^\alpha$ on the fibre:
\begin{equation}\label{eq:T3}
    \begin{array}{c}\vspace{.2cm}
    g= g_{ij}dr^i\cdot dr^j+ h_{\alpha \beta}(d \phi^\alpha+ \lambda^\alpha_i dr^i)\cdot(d \phi^\beta+ \lambda^\beta_j dr^j)\ , \\
    B=\frac12 b_{ij}d r^i \wedge d r^j+b_{\alpha i}dr^i\wedge
    (d \phi^\alpha+\frac12 \lambda^\alpha_j dr^j)+
    B_{\alpha \beta}(d \phi^\alpha+ \lambda^\alpha_i dr^i)\wedge(d \phi^\beta+ \lambda^\beta_j dr^j)\ ;
    \end{array}
\end{equation}
the $\lambda_i^\alpha$ are connections for the $T^3$--fibration.
Consider the action for example by an element of ${\rm O}(3,3)$
(since the fibre is 3--dimensional) of the form
\begin{equation}\label{eq:beta}
    \left(\begin{array}{cc}\vspace{.2cm}
        1 & 0\\
        \beta_{\rm T} & 1
    \end{array}\right)
\end{equation}
for $\beta_{\rm T}^{\alpha \beta}$ some {\it real} bivector on $T^3$. Then one can show that (taking the initial $B_{\alpha \beta}=0$
for simplicity)
the Buscher rules for this element can be summarized by acting
on (\ref{eq:T3}) by the simple rules
\begin{equation}
    \lambda^\alpha_i\to \lambda^\alpha_i+ \beta_{\rm T}^{\alpha \beta}b_{\beta i}\ , \qquad b_{\alpha i}\to b_{\alpha i}\ , \qquad
    (h+B)_{\alpha \beta}\to \left(h \frac 1{1+\beta h}\right)_{\alpha \beta}\ ;
\end{equation}
this result is very similar to the one in \cite{fmt} for the simpler inversion along the three direction of the fibre.

The result of the previous paragraph can be applied to $\cc^3$, viewed
as a $T^3$ fibration on $\rr_+^3$. We see that, since the $b_{\alpha i}$ are vanishing to begin with, the $\lambda_i^\alpha$ do not
change. Since the $\lambda_i^\alpha$ are the 
connections for the $T^3$--fibration, the
manifold stays topologically the same.

This leaves us with the easier part of the T--duality action. Since vectors and one--forms are mixed by an element of $O\in{\rm O}(6,6)$,
the pure spinor $\Phi$ is also acted on\footnote{$O$ also acts on the corresponding generalized complex structure ${\cal J}$ as $(O^t)^{-1}{\cal J}O^t$, see \cite[Eq.(6.9)]{gmpt3}. Applying this
to (\ref{eq:beta}), we obtain what is called $\beta$--transform in \cite[Ex.2.2]{gualtieri}.}. In general, an endomorphism of $T\oplus T^*$ also acts naturally on differential forms, as detailed for example in \cite[Section 2.1]{t-reform}. For the
case of interest here, (\ref{eq:beta}), this is nothing but the
action of the bivector $\beta_{\rm T}$ by contraction:
\begin{equation}\label{eq:betaPhi}
    \Phi_\pm = e^{\beta_{\rm T}\llcorner} \Phi_\pm^0\ .
\end{equation}
The bivector proposed in \cite{lunin-maldacena} is one that
exhibits $\zz_3$ symmetry, as (\ref{eq:wls}) does:
\begin{equation}\label{eq:betaphi}
    \beta_{\rm T}= 4\gamma(\iota_{\del_{\varphi^1}}\wedge\iota_{\del_{\varphi^2}}+{\rm cycl.\ perm.})\ .
\end{equation}
We emphasize once again that $\beta_{\rm T}$ is real, as opposed
to $\beta$ in section \ref{sub:moduli}, which was complex. (At the
first order level, in (\ref{eq:bivdef}), one can freely add to
$\beta$ its complex conjugate.)

If one acts with (\ref{eq:betaphi}) on the flat space pure spinors
$(\Phi_+^0,\Phi_-^0)=(e^{-i J_0},\Omega_0)$ as in  (\ref{eq:betaPhi}),
one obtains the LM--NS solution we promised. One can easily check,
for example, that
\begin{equation}\label{eq:betadw}
    \beta_{\rm T}\llcorner (dz^1 \wedge dz^2 \wedge dz^3) = d(\gamma \,z^1 z^2 z^3)\ ;
\end{equation}
comparing with (\ref{eq:wls}), we see that the right hand side
is exactly $dw$ for the case we restricted ourselves to, $\gamma'=0$.

In spite of this initial success, however, it turns out
that the rest of the pure spinor pair (\ref{eq:betaPhi}) is not of the form
(\ref{eq:Bdielpuretilde}). Since we advertised those as the most general pair of types 0  and 1, this would appear to be a problem.
Fortunately, the difference is explained by a simple change of
gauge in the $B$--field, as follows. In the language of \cite{minasian-petrini-zaffaroni}, the $B$--field obtained after the
action (\ref{eq:betaPhi}) is
\begin{equation}
    B_0= -\cos(2 \psi)\sin(2 \psi)y_1 \wedge y_2\ ;
\end{equation}
this differs from the one in (\ref{eq:B}) by
\begin{equation}
    \delta b=\tan(2 \psi) x_1 \wedge x_2=\frac14 d r_1^2\wedge
    d r_2^2+{\rm cycl.\ perm.}
\end{equation}
which is exact, and hence a gauge transformation.

A curious fact is that if one now adds this to the action of $\beta$
in (\ref{eq:betaPhi}), the flat--space $\Phi_+=e^{-iJ_0}$ is invariant
in form:
\begin{equation}\label{eq:emiJ}
    \Phi_+=e^{\beta_{\rm T}\llcorner + \delta b\wedge}
    e^{-i J_0}=e^{-i J_0}\ .
\end{equation}
(This does not mean that there is no $B$--field, since $B$ has to be
read off from the pair. $B$ is actually of the form (\ref{eq:B}).)
Indeed, one can see that the LM--NS solution is a particular case of the procedure (\ref{eq:fingKdef}). We suspect this feature
is general for duals to Leigh--Strassler theories, and some of the
perturbative results in \cite{aharony-kol-yankielowicz} seem to point in this direction. We were not able, however, to use this to find the solution.

Summarizing, the pure spinors of the LM--NS solution are given by
\begin{equation}
    \label{eq:lm}
    \Phi_\pm = \exp[\gamma(4\iota_{\del_{\varphi^1}}\wedge\iota_{\del_{\varphi^2}}+\frac14 d r_1^2\wedge
    d r_2^2)+{\rm cycl.\ perm.}]\wedge\Phi_{\pm0}
\end{equation}
where $(\Phi_{+0},\Phi_{-0})=(e^{-i J_0},\Omega)$ is the flat solution
and $z^i=r_i e^{i \varphi_i}$. Explicitly we have that
\begin{eqnarray}
\label{eq:lm+}\Phi_+&=&\Phi_{+0}=e^{-iJ_0}, \\
\label{eq:lm-}\Phi_-&=& dw\wedge
\exp[\frac{dz^1\wedge dz^2}{\gamma z^1 z^2}+
\frac \gamma4 (dr_1^2 \wedge dr_2^2+ {\rm cycl.\ perm.})] .
\end{eqnarray}
One prominent feature of this pair of pure spinors is that they are exact at second order\footnote{The deformations in \cite{lin-tolman,bursztyn-cavalcanti-gualtieri} are exact at first order; however, as mentioned
in section \ref{sub:moduli}, the pure spinors there are not closed.} in $\gamma$: the perturbation theory truncates. This is not true of the metric, which receives corrections at all orders. This interesting situation is due to the fact that passing from the pure spinors to the metric is non--linear; we regard this as encouragement that generalized complex geometry is the right framework to find more general solutions of this sort.


\subsection{Lunin--Maldacena from a potential} 
\label{sub:lm_pot}

We saw how the LM--NS solution is defined by the action of
(\ref{eq:betaphi}) as in (\ref{eq:betaPhi}), and that we can
gauge--transform it in such a way as to fall in the class (\ref{eq:Bdielpuretilde}).

Now we want to put this solution in the form (\ref{eq:pureK}). We have already noticed how the coordinate $w$ has to be taken
\begin{equation}
    \label{eq:wlm}w=\gamma \, z^1 z^2 z^3 \ .
\end{equation}
The semi-chiral co-ordinates are
\begin{equation}\label{eq:qplm}
\begin{array}{cc}\vspace{.2cm}
    q=\log(z^1)-\frac \gamma 2 (|z^2|^2-|z^3|^2)\ , \qquad
    p=\log(z^2)-\frac \gamma 2 (|z^3|^2-|z^1|^2)\ , \\
    Q=\log(z^1)+\frac \gamma 2 (|z^2|^2-|z^3|^2)\ , \qquad
    P=\log(z^2)+\frac \gamma 2 (|z^3|^2-|z^1|^2)\ ,
\end{array}
\end{equation}
which agrees with eq. (\ref{eq:Om6}), (\ref{eq:pureK}) and (\ref{eq:lm-}).

Moreover, if one chooses, as usual, $q, P, w$ and their complex coordinates as
independent variables, the generating function $K$ is
\begin{equation}
    \label{eq:Klm}
    K= q P + \bar q \bar P + \gamma(|z^1|^2+ |z^2|^2 + |z^3|^2)+
    \gamma^2(|z^2|^2-|z^3|^2)(|z^3|^2-|z^1|^2)
\end{equation}
where one has to understand $z^i=z^i(q,P,w,\bar q, \bar P, \bar w)$
given by inverting (\ref{eq:wlm}) and (\ref{eq:qplm}). (It is not
necessary to invert them explicitly. To check that  $K$ in (\ref{eq:Klm}) satisfies (\ref{eq:pQ}), one can compute the Jacobian $\frac{\del(q,P,w)}{\del(z^1,z^2,z^3)}$) and invert it.)

One can check that the equation (\ref{eq:det}) is satisfied for
this choice, and that 
$K$ is not just quadratic in $q,P,w$.

\bigskip \bigskip

{\bf Acknowledgments.} We would like to thank S.~Benvenuti, M.~Headrick, O.~Lunin, J.~Maldacena, D.~Martelli, I.~Melnikov,  M.~Mulligan, M.~Petrini, M.~Ro\v cek, E.~Witten, M.~Zabzine for discussions.
A.~T.~is supported by the DOE under contract DEAC03-76SF00515 and by the NSF under contract 0244728. N.~H.~ is supported in part by NSF Grants PHY-0094328 and PHY-0401814 and a Fermi-McCormick Fellowship.
Both authors would like to thank the Aspen Center for Physics for hospitality during the completion of this project.


\providecommand{\href}[2]{#2}

\end{document}